\documentstyle[prd,eqsecnum,aps,amsfonts]{revtex}

\begin{document}
\draft
\title{Just how long can you live in a black hole and what can be done
about it?}
\author{Gregory A. Burnett}
\address{Department of Physics, University of Florida,
Gainesville, Florida\ \ 32611}
\date{19 April 1995}
\maketitle
\begin{abstract}
We study the problem of how long a journey within a black hole can
last.  Based on our observations, we make two conjectures.  First, for
observers that have entered a black hole from an asymptotic region, we
conjecture that the length of their journey within is bounded by a
multiple of the future asymptotic ``size'' of the black hole, provided
the spacetime is globally hyperbolic and satisfies the dominant-energy
and non-negative-pressures conditions.  Second, for spacetimes with
${\Bbb R}^3$ Cauchy surfaces (or an appropriate generalization
thereof) and satisfying the dominant energy and non-negative-pressures
conditions, we conjecture that the length of a journey anywhere within
a black hole is again bounded, although here the bound requires a
knowledge of the initial data for the gravitational field on a Cauchy
surface.  We prove these conjectures in the spherically symmetric
case.  We also prove that there is an upper bound on the lifetimes of
observers lying ``deep within'' a black hole, provided the spacetime
satisfies the timelike-convergence condition and possesses a maximal
Cauchy surface.  Further, we investigate whether one can increase the
lifetime of an observer that has entered a black hole, e.g., by
throwing additional matter into the hole.  Lastly, in an appendix, we
prove that the surface area $A$ of the event horizon of a black hole
in a spherically symmetric spacetime with ADM mass $M_{\text{ADM}}$ is
always bounded by $A \le 16\pi M_{\text{ADM}}^2$, provided that future
null infinity is complete and the spacetime is globally hyperbolic and
satisfies the dominant-energy condition.
\end{abstract}
\twocolumn

\section{Introduction and Summary} \label{sec:intro}
Over the past forty years, black-hole physics has become quite a
mature field \cite{Thorne94,Israel89}.  Driven by a desire to
understand their fundamental properties and potential importance to
astrophysics, the study of black holes has produced a number of
outstanding results.  For example, Hawking has proven that black holes
never bifurcate and that the areas of black-hole event horizons never
decrease to the future \cite{HawkingEllis73,Wald84}.  Further, through
the efforts of a number of researchers, the black hole ``no-hair''
theorems \cite{Wald84,Chrusciel} have established that stationary
electrovacuum black holes are far simpler objects than were once
imagined \cite{Israel89,Carter73}.  In all of these studies, the
interiors of black holes have been entirely ignored, as well they
should as these regions neither affect nor are observable from the
asymptotic region of the spacetime (by their very definition).  So,
while a great deal has been learned about black-hole exteriors,
comparatively little is known about black-hole interiors.

Are there any features that black-hole interiors share?  There are at
least two results in this direction.  First, Penrose has shown that if
a spacetime contains a future trapped surface then, provided the
cosmic censorship hypothesis holds
\cite{Penrose69,Penrose78,Penrose79,GerochHorowitz79,Wald84}, it must
lie within the black-hole region and that such a spacetime must be
singular in the sense that it cannot be future null geodesically
complete \cite{Penrose65,HawkingEllis73,Wald84}.  While a truly
amazing and important theorem in the history of black-hole physics, it
tells us very little about the extent to which the spacetime is
singular.  Are just a few, most, or all of the geodesics in the
interior region incomplete?  How long does it take for such a
singularity to develop?  Second, a number of people using a variety of
methods (see \cite{PoissonIsrael90} and the references therein), have
examined whether the Cauchy horizons that occur in the interiors of
the Kerr-Newman black holes will be present in more ``realistic''
spacetimes.  In particular, Poisson and Israel have argued that these
horizons are destroyed (generically) due to an unbounded ``inflation''
of the mass function $m$ on what was the Cauchy horizon inside the
black hole \cite{PoissonIsrael90,BIP90}.  Although arguments have been
given that indicate that these results are quite general, a proof
remains to be found \cite{BDIM94a,BDIM94b}.

Here, in the effort to explore the features of black-hole interiors,
we examine a single issue: How long can the journey of an observer in
a black hole last?  Must it be finite?  If so, can an upper bound on
the length of such a journey be given in terms of some characteristic
of the black hole?  It turns out that there are two forms of this
problem that are best considered separately.  First, we restrict
ourselves with observers that enter a black hole from the asymptotic
region.  That is, we are restricting our attention to the portion of a
black-hole region lying to the future of past null infinity ${\cal
J}^-$.  Second, we drop this restriction and look at the entire
black-hole region.  However, we shall see that this generalization
does come at a small price.  A restriction on the Cauchy surface
topology is necessary (while none appears necessary for the first
case) along with a weaker bound on the lifetimes of observers within
the black-hole region.

\begin{figure}
\input{psfig}
\centerline{\psfig{figure=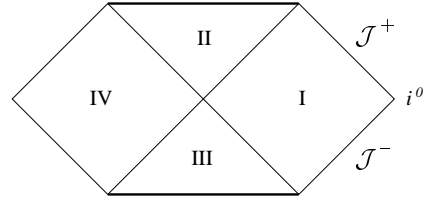}}
\medskip
\caption{A spacetime diagram representing a maximally extended
Schwarzschild spacetime of positive mass $M$.  The black-hole region
(according to ${\cal J}^+$), is all of regions II and IV.  An observer
beginning a journey in the asymptotic region $\langle\langle{\cal
J}\rangle\rangle$ (region I) and entering the black hole must enter
region II.  A journey in this region cannot last more than a time
$\pi M$ (and this maximum is attained for journeys perpendicular to
the surfaces of spatial homogeneity).}
\label{schwarz1}
\end{figure}

Consider first the case of an observer entering a black hole from an
asymptotic region \cite{af_defs}
\begin{equation}
\langle\langle{\cal J}\rangle\rangle = 
J^-({\cal J}^+) \cap J^+({\cal J}^-),
\end{equation}
also known as the domain of outer communications \cite{Carter73}.  In
the maximally extended Schwarzschild spacetime, this is region II
depicted in Fig.~\ref{schwarz1}.  A partial answer to our problem can
obtained from the work of Friedman, Schleich, and Witt which can be
stated as: In an asymptotically flat spacetime satisfying the
null-convergence condition \cite{ec} with distinct null infinities
${\cal J}_i$, the sets $J^-({\cal J}^+_i)
\cap J^+({\cal J}^-_j)$ are empty for $i \neq j$ \cite{FSW93}.
Therefore, fixing one null infinity ${\cal J}_0$ it is impossible for
an observer to leave the asymptotic region $\langle\langle{\cal
J}_0\rangle\rangle$ by entering a black hole and then live
indefinitely by escaping to another asymptotic region.  The maximally
extended Schwarzschild spacetime depicted in Fig.~\ref{schwarz1}
exemplifies this impossibility.  Although this result is very
encouraging, it does not eliminate other possibilities for an observer
existing indefinitely within a black hole.  Based on the general
considerations presented in Sec.~\ref{sec:formulating} and its
successful proof in the spherically symmetric case (theorem~1 below),
we conjecture that this result can indeed be strengthened.

{\it Conjecture 1.} Fix a globally hyperbolic asymptotically flat
spacetime satisfying the dominant-energy and non-negative-pressures
conditions \cite{ec} with a non-empty black-hole region.  Fix a
connected component ${\cal E}_0$ of the event horizon and denote the
supremum of its area as measured over the set of all of its spatial
cross-sections by $A_f$.  Then, there exists a constant $k$
(independent of which spacetime and event horizon is chosen) such that
the lifetime of any observer within the black hole is bounded above by
$k
\sqrt{A_f/16\pi}$ provided that the black hole is entered by crossing
${\cal E}_0$ in the future of ${\cal J}^-$.

Recall that the irreducible mass associated with a connected component
of black-hole's event horizon at a given ``time'' is obtained by
calculating its surface area $A$ on a spatial cross-section and
setting \cite{Wald84}
\begin{equation} \label{def_M_irr}
M_{\text{irr}} = \sqrt{A/16\pi}.
\end{equation}
The quantity is ``irreducible'' as it is non-decreasing to future,
which follows from the fact that the area $A$ of the event horizon is
non-decreasing to the future \cite{HawkingEllis73,Wald84}.  So, with
this definition, conjecture~1 bounds the lifetime of observers within
a black hole by a multiple of the supremum of the irreducible
mass calculated on the black-hole's event horizon (provided the black
hole is entered from the asymptotic region).

To explore the features of conjecture~1, consider a black-hole
spacetime associated with two distant stars that each collapses
forming black holes.  Should these two black holes never collide with
one another, the event horizon will then have two connected components
and conjecture~1 asserts that the lifetime of an observer that enters
one of the black holes (to the future of ${\cal J}^-$) is bounded by a
quantity computable from the geometry of the event horizon of the
black hole entered.  However, should the two black holes collide and
coalesce into a single black hole, the event horizon will then have a
single connected component.  In this case, the bound given by
conjecture~1 is insensitive as to whether the black hole was entered
before or after coalescence and to which black hole was entered.  So,
the bound is the same for observers entering a small black hole and a
large black hole as long as the two eventually coalesce.  This is
probably a gross overestimate in the case of an observer entering the
smaller black hole before coalescence and probably can be refined.

Next, suppose we slightly modify conjecture~1 by having it place a
bound on the lifetime of observers in the portion of the black hole
lying to the future of ${\cal J}^-$.  Fix any two points $p$ and $q$
in this region with $q \in J^+(p)$ and fix any causal curve $\nu$ from
${\cal J}^-$ to $p$.  Then, as $\nu$ must intersect the event horizon
somewhere, conjecture~1 gives an upper bound on the lengths of all
causal curves from $p$ to $q$ (that is independent of $p$ of $q$),
thereby bounding the lifetimes of all observers in this region.
However, suppose that the black hole is sufficiently complicated that
its boundary (part if not all of the total event horizon) is not
connected.  For example, we could imagine a spacetime containing a
wormhole with its mouths sufficiently separated and moving in such a
way that the event horizons enclosing each mouth never coalesce.
Then, if $p$ can be reached from ${\cal J}^-$ by two causal curves
that intersect different horizons (i.e., thread different mouths), we
would have a choice of which component to use in establishing an upper
bound to the lifetimes of observers therein.  Can we use the smaller
of the two, or must we always use the larger?  Or should we use a
bound based on total area of the event horizons associated with our
region?  We leave this problem open for investigation.

In conjecture~1, we have limited ourselves to observers entering a
black hole from an asymptotic region.  Might the bound given by
conjecture~1 hold for all observers within a given black hole?
Clearly not, if we do not put some restriction on the Cauchy surface
topology as the maximally extended Schwarzschild spacetime provides a
simple counterexample.  Here, the problem is that there as an entire
asymptotic region within the black hole.  (See Fig.~\ref{schwarz1}.)\
\  Even if we restrict our consideration to spacetimes with Cauchy
surfaces having the topology of a compact orientable connected
three-manifold $C$ minus a point (so that there are no other
asymptotic regions to escape to) then, as is demonstrated in
Sec.~\ref{sec:topology}, if we do not further restrict $C$,
asymptotically flat black-hole spacetimes can be constructed with
``compact internal regions'' and yet admitting infinitely long
timelike curves in the black-hole region.  Furthermore, as is
demonstrated in Sec.~\ref{sec:formulating}, even if we restrict
ourselves to spacetimes with ${\Bbb R}^3$ Cauchy surfaces, there is no
$k$ for which the bound given by conjecture~1 will hold.  However,
here the only problem appears to be that the bound is not quite
appropriate.  The idea is that the bound in conjecture~1 is
constructed from the geometry of the event horizon and therefore can
be calculated without a knowledge of the interior of the black hole.
For more general observers, a knowledge of the interior geometry (at
least on a Cauchy surface) is necessary in obtaining a bound on the
lifetimes of observers therein.  Based on these considerations and its
successful proof in the spherically symmetric case (theorem~2 below),
we are led to the following conjecture.

{\it Conjecture 2.}  Fix a globally hyperbolic asymptotically flat
spacetime satisfying the dominant-energy and non-negative-pressures
conditions \cite{ec}.  Then, provided the Cauchy surfaces are
diffeomorphic to ${\Bbb R}^3$, there is a number $k$ (independent of
which spacetime is chosen) and a quantity $M_\Sigma$ computable from
the initial data on a Cauchy surface $\Sigma$, such that the lifetimes
of all observers within the black-hole region is bounded above by $k
M_\Sigma$.

In this conjecture, $M_\Sigma$ is intended to represent the least
upper bound to a quasilocal mass function (associating a number with
each given spacelike two-sphere in a spacetime) over the set of all
two-spheres in $\Sigma$.  As we have not specified what quasilocal
mass function is to be used (see however Ref.~\cite{Hayward94} and the
references therein), this portion of the conjecture is imprecise and
we leave the task of making it precise as part of the problem.
However, when we specialize to the spherically symmetric spacetimes,
we expect this mass to be simply $\sup_\Sigma(m)$, where $m$ is the
mass function constructible in any spherically symmetric spacetime as
given by Eq.~(\ref{defm}) in Sec.~\ref{sec:elementaries}.

For simplicity, we have limited the Cauchy surface topology in
conjecture~2 to ${\Bbb R}^3$.  In Sec.~\ref{sec:topology}, we
generalize this conjecture by broadening the class of spatial
topologies.  Furthermore, a theorem is presented which states
(roughly) that the lengths of causal curves in the portion of the
black hole not reachable from the asymptotic region are bounded
provided the spacetime admits a maximal Cauchy surface.  Although this
is far from a proof of conjecture~2, it is encouraging.

As the calculation of $A_f$ in conjecture~1 requires a knowledge of
the geometry of spacetime in the far future (and therefore can be
quite difficult to calculate in practice) one might worry that $A_f$
could be arbitrarily large (or even infinite) thereby making the bound
given by conjecture~1 rather weak.  However, from an analysis of the
spherically symmetric case (see theorem A1 in Appendix~A), we expect
the three masses $M_{\text{irr}}$, $M_{\text{ADM}}$, and $M_\Sigma$ to
be related according to the following inequalities
\begin{equation} \label{mass_ineq}
M_{\text{irr}} \le M_{\text{ADM}} \le M_\Sigma.
\end{equation}
Whether the second inequality actually holds cannot be investigated
until $M_\Sigma$ is given a precise definition.  However, since the
total area $A_T$ of the black hole at a given ``time'' is the sum of
the areas associated with each connected component, the first
inequality always holds provided that $A_T \le 16\pi
M_{\text{ADM}}^2$.  (This inequality is similar to, though distinct
from, the Gibbons-Penrose isoperimetric inequality for initial data
sets in general relativity which has its inspiration from this
inequality \cite{Gibbons72,Penrose73,JangWald77,Jang79,Gibbons84}.)\ \
That such an inequality should hold can be argued as follows.
Consider this inequality in an asymptotically flat electrovacuum
spacetime whose event horizon is connected (i.e., so there is only one
black hole at late times).  As the area of the event horizon is
non-decreasing to the future, this inequality is most difficult to
satisfy at late times.  However, it is expected that the black hole
will eventually settle-down and approach a stationary state which, by
the ``no-hair'' theorems, must be one of the Kerr-Newman spacetimes
where it is known that the inequality does indeed hold.  (As far as
the author is aware, whether this inequality always holds is
unresolved.  Theorem~A1 in Appendix~A proves it does hold for the
spherically symmetric spacetime satisfying the dominant-energy
condition.)  Therefore, should the first inequality in
Eq.~(\ref{mass_ineq}) indeed hold, then the bound given by
conjecture~1 will always be finite.  Furthermore, independent of this
inequality, an alternate version of conjecture~1 can be obtained by
replacing the bound by $k M_{\text{ADM}}$.  Should the irreducible
mass always be bounded by the ADM mass (as expected), then
conjecture~1 would imply this alternate version.

\begin{figure}
\input{psfig}
\centerline{\psfig{figure=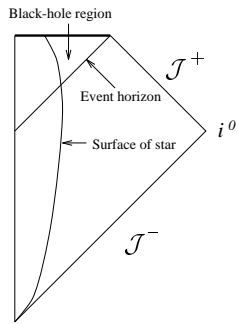}}
\medskip
\caption{A spacetime diagram representing the collapse of a star
forming a black hole.  Notice that, in this simple case, the event
horizon can be used as a Cauchy surface for the black-hole region.}
\label{collapse}
\end{figure}

Conjectures~1 and~2 are similar to the conjecture that closed
universes possessing $S^3$ or $S^1 \times S^2$ Cauchy surfaces (and
obeying the dominant-energy and non-negative-pressures conditions)
must have finite lifetimes
\cite{Burnett91,Burnett93,Burnett94,Burnett95}, which is a slight
variant of the closed-universe recollapse conjecture
\cite{BGT85,BarrowTipler85,BarrowTipler88,MarsdenTipler80}.
In fact, the similarity is more than superficial.  The idea is that
the requirement that the Cauchy surfaces be compact in the
closed-universe finite lifetime conjecture can probably be replaced by
the more general requirement that the geometry on Cauchy surfaces be
bounded in an appropriate sense.  In the spherically symmetric case we
can make this precise.  If we replace the requirement that the Cauchy
surfaces be compact with bounds on $r$ and $m$, we can again establish
a finite lifetime for the spacetime.  (Actually, if this is done, the
final bound on the lifetime may be slightly different from that stated
in Ref.~\cite{Burnett95}.)\ \ With this in mind, consider the collapse
of a star leading to the formation of a black hole as depicted in
Fig.~\ref{collapse}.  The black-hole region $B$ of this spacetime is
globally hyperbolic with its boundary ${\cal E} = \partial B$ (the
event horizon of the black hole) acting as a Cauchy surface for this
region.  Although ${\cal E}$ is non-compact it does enjoy the property
that the geometry thereon is essentially bounded.  In particular, the
area of the black hole (as measured on the event horizon) is bounded
by its future asymptotic value.  Therefore, it would seem that proofs
of these two conjectures would be very similar and that one could be
obtained from the other with only slight modifications being required.
Indeed, a proof of conjectures~1~and~2 for the spherically symmetric
case can be obtained using the techniques developed in the course of
the proof that spherically symmetric closed universes have finite
lifetimes.  In fact, we prove somewhat more.

{\it Theorem 1.} Fix a globally hyperbolic asymptotically flat
spherically symmetric spacetime satisfying the dominant-energy and
non-negative-pressures conditions \cite{ec} and possessing a
Cauchy surface that is geodesically complete.  Then, the lifetime of
an observer in the black hole is bounded above by $20~\sup_{\cal
E}(M_{\text{irr}})$ provided either: (a) the black hole is entered
from the asymptotic region of the spacetime; or (b) the spacetime
possesses a spherically symmetric Cauchy surface with $\nabla^a r$
everywhere outgoing; or (c) the observer lies to the future of the
component of a Cauchy surface connected to $i^0$ on which $\nabla^a r$
is outgoing.

Note that in the spherically symmetric case, on a sphere of symmetry
of area $A = 4\pi r^2$, $M_{\text{irr}}$ is simply $r/2$.
As shown by theorem~A1 in Appendix~A, provided the dominant-energy
condition holds, for spheres on the event horizon $M_{\text{irr}} \le
M_{\text{ADM}}$ (i.e., $r/2$ on the event horizon is always bounded
above by $M_{\text{ADM}}$).  Therefore, the bound in theorem~1 can
always be replaced by $20~M_{\text{ADM}}$ thereby given a bound that
is calculable from the initial data on a Cauchy surface.  Further,
provided that the event horizon is complete in the sense given by
theorem~A2 in Appendix~A, then $M_{\text{irr}}=r/2$ and $m$ have the
same future asymptotic limits on the event horizon.  Therefore, in
this case, the bound in theorem~1 is equivalent to $20$ times the
future asymptotic value of the mass $m$ on the event horizon.

Each part of theorem~1 has its own strengths and weaknesses.  Part (a)
of this theorem is best used when we know that the observer is
entering the black hole from the asymptotic region.  However, suppose
we are given the location of an observer on a (spherically symmetric)
initial data surface $\Sigma$ (a Cauchy surface) and that he enters
the black hole.  Do we need to find the rest of the spacetime from the
initial data on $\Sigma$ to find out whether the observer was indeed
in the asymptotic region?  No.  If $\nabla^a r$ is outgoing on all of
$\Sigma$, then by (b) the length of his (and indeed all observers)
journey in the black hole is bounded by a multiple of the future
asymptotic irreducible mass of the black hole.  However, suppose that
$\nabla^a r$ is not outgoing on all of $\Sigma$ (i.e., $\Sigma$
contains a future or past outer trapped surface).  Then, provided our
observer begins his journey on the portion of $\Sigma$ connected to
$i^0$ on which $\nabla^a r$ is outgoing, then, by (c), his lifetime
within the black hole is still bounded by a multiple of the future
asymptotic irreducible mass of the black hole.

In theorem~1, we have demanded that the spacetime admit a Cauchy
surface that is geodesically complete.  This is automatically
satisfied by asymptotically flat Cauchy surfaces with ${\Bbb R}^3$
topology, but not may not be satisfied by those with ${\Bbb R} \times
S^2$ topology.  We make this requirement as it guarantees that $m$
will be everywhere non-negative, which is needed for the method of
proof used herein (see lemma~1).  It is likely, though not for
certain, that the proof can be improved upon, allowing us to lift the
requirement that a Cauchy surface be complete.

The next theorem gives an upper bound on the lifetimes of observers
within the black-hole region of a spherically symmetric spacetime that
makes no requirements on how the black hole was entered nor on how
$\nabla^a r$ behaves on a Cauchy surface.  Note that the price for
this is that a restriction on the Cauchy topology is necessary and
that the bound need not be as tight as that given by theorem~1.

{\it Theorem 2.} Fix an asymptotically flat spherically symmetric
spacetime that possesses ${\Bbb R}^3$ Cauchy surfaces and that
satisfies the dominant-energy and non-negative-pressures conditions
\cite{ec}.  Then, for any Cauchy surface $\Sigma$, the lifetime of an
observer in the black-hole region $B$ with event horizon ${\cal E} =
\partial B$ is bounded above by
\begin{equation} \label{theorem2}
20~\max \biglb( \sup_{{\cal E}}(M_{\text{irr}}),
\max_{(\Sigma \cap B) \cup P}(m) \bigrb),
\end{equation}
where $P$ is the subset of $\Sigma$ on which $\nabla^a r$ is
past-directed timelike, past-directed null, or zero.

To understand the features of the bound given by theorem~2, consider
the black hole formed by the collapse of a spherical star from a state
that is benign in the sense that on a spherically symmetric Cauchy
surface $\Sigma$ there are no future or past outer trapped surfaces
(so in particular $\nabla^a r$ is outgoing everywhere thereon).  Then,
$P$ is empty, the maximum of $m$ on $\Sigma \cap B$ occurs on its
boundary (which lies on ${\cal E}$) where it is no greater than
$M_{\text{irr}}$.  So, in this case, theorem~2 reproduces the bound
given by part (b) of theorem~1.  If, however, the spacetime contains
no such Cauchy surface (see for example the black-hole spacetime given
by Fig.~\ref{db2}), then $\max_{(\Sigma \cap B) \cup P}(m)$ can be
larger than $\sup_{\cal E}(M_{\text{irr}})$.  As discussed in
Sec.~\ref{sec:formulating}, this larger bound is necessary.
Furthermore, as $(\Sigma \cap B) \cup P$ is a subset of
$\Sigma\setminus \langle\langle{\cal J}\rangle\rangle$, this quantity
is constructed from the geometry on the portion of $\Sigma$ that is
not in the asymptotic region.  Lastly, we note that the rather
complicated bound given by theorem~2 can always be replaced by the
simpler but cruder bound $20~\sup_\Sigma(m)$ as $(\Sigma \cap B)\cup
P$ is a subset of $\Sigma$ and $\sup_{{\cal E}}(M_{\text{irr}}) \le
M_{\text{ADM}} \le \sup_\Sigma(m)$.  (Note that this theorem suggests
a strengthening of conjecture~2.)

The bound given by theorem~1 has the feature that it increases as
matter flows through the event horizon into the black hole thereby
making it larger.  Therefore, it might seem that we could help an
observer that has entered a black hole live longer by gathering up all
the material available to us outside the black hole and throw it in
after him.  Does increasing the black-hole's mass in this way really
help the observer?  In Sec.~\ref{sec:do}, we formulate and study a
precise version of this problem.  We find that while generally the
lifetime of the observer can be increased, there are circumstances in
which our effectiveness is limited and others in which there is
nothing that can be done to help the unlucky traveler.

The remainder of this work is organized as follows.  In
Sec.~\ref{sec:formulating}, we go through the steps leading to the
formulation of conjectures~1 and~2.  In Sec.~\ref{sec:topology}, we
generalize the class of Cauchy surface topologies for which we believe
conjecture~2 holds and prove a theorem which bounds the lengths of
causal curves lying ``deep within'' a black hole in a spacetime
possessing a maximal Cauchy surface.  In Sec.~\ref{sec:spherical}, we
provide the proofs of theorems~1 and~2.  In Sec.~\ref{sec:do}, we
investigate the problem of whether we can increase the lifetime of an
observer within a black hole.  Lastly, in Sec.~\ref{sec:discussion},
we make a few final remarks and discuss the hopes for extending the
results presented.

Our conventions are those of Ref.~\cite{Wald84}.  In particular,
metrics are such that timelike vectors have negative norms and the
Riemann and Ricci tensors are defined by $2 \nabla_{[a}\nabla_{b]}
\omega_c = R_{abc}{}^d \omega_d$ and $R_{ab} = R_{amb}{}^m$,
respectively.  All metrics are taken to be $C^2$.  For two sets $A$
and $B$, $A \setminus B$ denotes the elements in $A$ that are not in
$B$ (i.e., the set difference), $A^c$ denotes the elements that are
not in $A$ (i.e., its complement), and $\partial A$ denotes the
boundary of $A$.  Lastly, our units are such that $G = c = 1$.

\section{Formulating and testing conjectures 1 and 2}\label{sec:formulating}
An observer beginning a journey from the asymptotic region of a
maximally extended Schwarzschild spacetime of positive mass $M$
(region I in Fig.~\ref{schwarz1}) into the black hole (regions II and
IV) must enter region II.  As is well known, this journey within the
black hole must end in a time no greater than $\pi M$ as the observer
will be crushed as he approaches the singularity within.  Does a
similar result hold more generally?  Consider the following
conjecture.

{\it Conjecture 3 (False).} There is a constant $k$ such that the
lifetime of any observer in a black hole of mass $M$ is no greater
than $kM$.

We begin by testing this conjecture against the stationary
(electrovacuum) black-hole spacetimes.  For a Kerr-Newman black hole
of mass $M$, angular momentum $Ma$, and electric charge $e$, with $M^2
> a^2 + e^2 > 0$, we immediately realize that some care is needed in
addressing this problem.  Here, an observer can travel into the black
hole and cross the Cauchy horizon associated with a Cauchy surface
$\Sigma$ that extends to spacelike infinity at both ends.  (See
Fig.~28(i) of Ref.~\cite{HawkingEllis73} or Fig.~12.4 of
Ref.~\cite{Wald84}.)\ \ After crossing this horizon, he can travel to
another asymptotically flat region and thereby live an indefinitely
long time.  However, that such a journey can be achieved is perhaps
not surprising once we realize that the observer, having gone outside
of the domain of dependence of $\Sigma$, is in a region that is
unpredictable (from the initial data on $\Sigma$).  It has been
conjectured (the cosmic censorship conjecture
\cite{Penrose69,Penrose78,Penrose79,GerochHorowitz79,Wald84}) and
investigations have indicated that such Cauchy horizons do not occur
for ``realistic'' spacetimes \cite{PoissonIsrael90,BIP90}.  So, while
a few spacetimes will have such horizons beyond which the journey of
an observer could continue (such as Kerr-Newman with $M^2 > a^2+e^2 >
0$) this will usually not be the case as the spacetime will become
sufficiently singular that the journey will be inextendible.  (Just
how singular it will become has become a matter of debate
\cite{BDIM94a,BDIM94b}.)  It is because of such counterexamples that
we have restricted ourselves to globally hyperbolic spacetimes in
conjectures~1 and~2.  Should either conjecture hold, we then have an
upper bound on the time that an observer can remain within the domain
of dependence of $\Sigma$ once inside the black hole.  Whether the
spacetime can be extended (so that the journey of the observer can
continue) or otherwise (so that the journey is at an end) is then a
matter for further investigation.

Restricting ourselves to the globally hyperbolic portion of the
Kerr-Newman spacetimes (with Cauchy surfaces extending to spacelike
infinity in both asymptotic regions) we find that conjecture~3 holds
with $M$ being the irreducible mass associated with the event horizon
(which is constant thereon) and $k \ge 2\pi$.  (Our reasons for
identifying $M$ with $M_{\text{irr}}$ and not $M_{\text{ADM}}$ or some
other quantity will be given below.)\ \ Therefore, the Kerr-Newman
spacetimes obey conjecture~1 provided $k \ge 2\pi$.  Furthermore,
these spacetimes show that $2\pi$ is the least value of $k$ for which
the conjecture can hold as it can be shown that the ratio of the
maximum lifetime of an observer within the portion of the black hole
lying to the future of ${\cal J^-}$ and $M_{\text{irr}}$, while always
less than $2\pi$, approaches $2\pi$ in a sequence of Kerr-Newman black
holes where $M^2 - a^2 - e^2 \rightarrow 0$ (i.e., the ratio
approaches $2\pi$ in the limit that the black hole becomes extreme).

In conjecture~3, no conditions are stated on the material content of
the spacetime.  Without such a condition we can easily counterexample
the conjecture as follows.  From an extended Schwarzschild spacetime
of positive mass, construct a new spacetime by multiplying the
Schwarzschild metric by a conformal factor that is unity outside the
black-hole region (i.e. on regions I and III in Fig.~\ref{schwarz1}).
Then, as the two spacetimes share the same causal structure and agree
outside the black-hole region of Schwarzschild, the share the same
black-hole region.  By choosing the conformal factor appropriately, we
can quite easily ensure the existence of arbitrarily long timelike
curves in the black-hole region (in particular, region II).  We
conjecture that spacetimes with ``ordinary'' matter do not exhibit
this type of behavior.  Just what restrictions on the matter content
is appropriate is an issue for investigation, however, in
conjectures~1 and~2 we have imposed the dominant-energy and
non-negative-pressures conditions \cite{ec}.  As is demonstrated by
theorems~1 and~2, together these conditions are sufficient for
establishing a theorem in the spherically symmetric case.  It is
likely that these conditions can be weakened as the Kerr-Newman
spacetimes satisfy the conclusions of conjecture~1 (with $k=2\pi$)
without satisfying the non-negative-pressures condition (in the $e
\neq 0$ case).

Further, in conjecture~3 there is no restriction on how the observer
enters the black hole.  This is important as the black-hole region
associated with the asymptotic region I of Schwarzschild is all of
regions II and IV in Fig.~\ref{schwarz1}.  In particular the other
asymptotic end is inside the black hole (with our definition of black
hole).  So, clearly there are infinitely long lived observers therein.
We need some type of condition to rule out this and similar
counterexamples.  Is it enough to restrict ourselves to observers that
begin their journey from outside the black hole?  No.  Consider an
observer that begins his journey in region III.  While initially
outside the black hole, he can go inside and reach the other
asymptotic region and thereby live forever.  However, an observer that
begins in the asymptotic portion, region I, must enter region II upon
entering the black hole wherein his lifetime must end in a finite
time.  So, one way to take care of this problem is to restrict
ourselves to observers who enter the black hole from ``sufficiently
far away''.  For example, we can restrict ourselves to observers that
begin their journey somewhere in the asymptotic region
$\langle\langle{\cal J}\rangle\rangle$ (region I in
Fig.~\ref{schwarz1}) as we have done in conjecture~1.  Or, somewhat
more weakly, we can restrict ourselves to observers that lie to the
future of ${\cal J^-}$.  A second possibility is to restrict ourselves
to spacetimes with ${\Bbb R}^3$ Cauchy surfaces as we have done in
conjecture~2.  This way, any possibility of escaping to another
asymptotic region is eliminated {\it ab initio}.  (More general
topologies are considered in Sec.~\ref{sec:topology}.)

\begin{figure}
\input{psfig}
\centerline{\psfig{figure=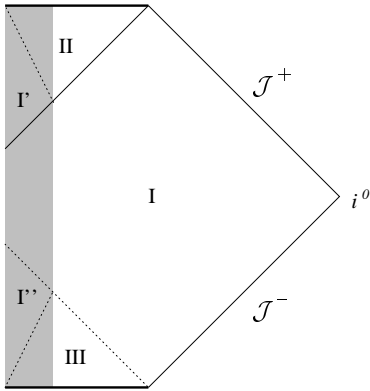}}
\medskip
\caption{A dust-ball spacetime constructed from a $k_{\text{RW}}=+1$
Robertson-Walker spacetime with $0 < \chi_0 < \pi/3$.  The black-hole
region here is the union of regions $\text{I}'$ and II.}
\label{db1}
\end{figure}

Lastly, in conjecture~3, we have not specified what is meant by the
mass $M$.  Should we use the irreducible mass (as we have done in
conjecture~1), the ADM mass of the spacetime, a quasilocal mass
constructed on the event horizon of the black hole, a quantity
constructed from the initial data (as we have done in conjecture~2),
or something else?  To investigate this issue, consider the black-hole
spacetimes associated with the gravitational collapse of uniform
density ``dust ball'' such as that depicted in Fig.~\ref{db1}.  (These
are the spacetimes considered by Oppenheimer and Snyder in their
pioneering work on gravitational collapse \cite{OppenheimerSnyder39}.)
These dust-ball spacetimes are easily constructed by ``gluing'' a
portion of a dust-filled Robertson-Walker spacetime to Schwarzschild
\cite{MTW73} and we refer the reader to Appendix~B for a brief review
of their construction and the establishment of notation for the
following.

\begin{figure}
\input{psfig}
\centerline{\psfig{figure=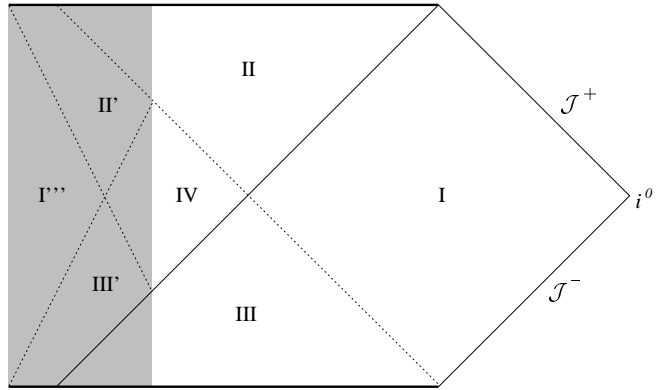}}
\medskip
\caption{A dust-ball spacetime constructed from a $k_{\text{RW}}=+1$
Robertson-Walker spacetime with $2\pi/3 < \chi_0 < \pi$.  The black
hole here is the union of regions $\text{I}'''$, II, $\text{II}'$,
$\text{III}'$, and IV.  An observer that begins his journey in region
I and enters the black hole, must enter region II.  On the other hand,
an observer beginning his journey from region III has the freedom to
visit all of the black-hole regions.  Notice that in this case every
Cauchy surface must intersect the black hole and that no spherically
symmetric Cauchy surface exists on which $\nabla^a r$ is outgoing
everywhere thereon.}
\label{db2}
\end{figure}

Note that for these spacetimes, the mass function $m$ coincides with
the ADM mass in the vacuum region and since all of the matter
eventually goes into the black hole, $M_{\text{irr}}$ and $m$ both
have $M_{\text{ADM}}$ as future asymptotic limits on the event
horizon.  So, while these spacetimes cannot distinguish which of these
three quantities should be used in conjecture~3, they can test whether
one (and hence all) will fail to work.  To see why all three are
inappropriate for conjecture~2, fix a dust-ball spacetime constructed
from a $k_{\text{RW}}=+1$ Robertson-Walker spacetime (so the dust ball
is gravitational bound) that is described by the constants $C$ and
$\chi_0 > 2\pi/3$.  (See Fig.~\ref{db2}.)\ \ Here, the black hole is
the union of regions $\text{I}'''$, II, $\text{II}'$, $\text{III}'$,
and IV.

For observers within the black hole, not necessarily entering from
region I, it can be argued that the maximum lifetime that can be spent
within the black hole is attained by the curve corresponding to the
center of the dust ball.  Therefore,
\begin{mathletters}
\begin{eqnarray}
\text{(maximum lifetime)} & = & \int_0^{2\pi} a(\eta) \; d\eta \\
                          & = & \pi C \label{life_C} \\
                          & = & \left( {2 \pi \over {\sin^3\!\chi_0}}
                                \right) M_{\text{ADM}} \label{coeff}.
\end{eqnarray}
\end{mathletters}
Now consider a sequence of such spacetimes in which $\chi_0
\rightarrow \pi$.  In this limit, the coefficient of $M_{\text{ADM}}$ 
in Eq.~(\ref{coeff}) diverges.  Therefore, this sequence of spacetimes
shows that, provided that we always identify $M$ as the ADM mass of
the spacetime (or the future asymptotic values of $M_{\text{irr}}$ or
$m$ on the event horizon), there is no constant $k$ for which the
above conjecture will hold.  This is not surprising once we realize
that in such a sequence (with fixed $C$) the maximum lifetime of
observers inside the black hole, given by Eq.~(\ref{life_C}), is
fixed, while the mass of the black hole, as measured using the ADM
mass given by Eq.~(\ref{dust-ball-adm-mass}) is becoming arbitrarily
small in this limit.  This shows, in particular, that an upper bound
on the lifetime of observers inside of a black hole cannot, in
general, be determined by examining the spacetime outside of the black
hole.  With $\chi_0$ close to $\pi$, the size of the spheres of
symmetry inside the black hole and the amount of matter therein can be
much greater than would be guessed by examining the mass of the black
hole from the outside.  Indeed, the spacetime within the black hole
acts very much like a closed universe (in particular like a
$k_{\text{RW}}=+1$ Robertson-Walker spacetime) except that it is
connected to an asymptotically flat spacetime by a small ``throat''.

So, conjecture~2 would fail if we identified $M_\Sigma$ with
$M_{\text{ADM}}$ or the future asymptotic values of $M_{\text{irr}}$
or $m$ on the event horizon.  Is there some other quantity that we can
construct that is sensitive to the geometry inside of the black hole?
Note that for these spacetimes with $k_{\text{RW}}=+1$ and $\chi_0 >
\pi/2$, the mass $m$ is not maximal on the outer boundary of the dust
ball.  It is maximal on the surface $\chi=\pi/2$.  Fixing a
spherically symmetric Cauchy surface $\Sigma$ in our spacetime and
using Eq.~(\ref{m_Robertson-Walker}), we see that the supremum of $m$
over $\Sigma$ is simply $C/2$.  Therefore, by Eq.~(\ref{life_C}), for
these spacetimes we have
\begin{equation}
\text{(maximum lifetime)} =  2\pi \sup_\Sigma(m).
\end{equation}
So, conjecture~2 does hold for these spacetimes provided that we take
$M_\Sigma$ to be the supremum of $m$ over a Cauchy surface and $k \ge
2\pi$.  Analyzing the remaining dust-ball spacetimes, we again find
that conjecture~2 holds provided that $k \ge 4+\pi$ and $M_\Sigma =
\sup_\Sigma(m)$.  (Note that this lower bound on the allowed values of
$k$ is not least.  Finding the least upper bound to the lengths of
timelike curves in the black holes of the dust-ball spacetimes
with $k_{\text{RW}}=-1$ or $k_{\text{RW}}=+1$ and $\chi_0 < \pi/2$
does not appear to be an easy task.  For instance, is the geodesic at
the center of the dust ball longest?  With some work it can probably
be shown that the conjecture holds for all of these spacetimes with $k
\ge 2\pi$.)

From these observations and in view of theorem~2, it would seem that
the proper identification for $M_\Sigma$ conjecture~2 would be
something that generalizes $\sup_\Sigma(m)$.  The most obvious
candidate is to generalize $m$ by some quasilocal mass function and
then find its supremum over the set of all two-spheres on a Cauchy
surface $\Sigma$.  However, whether this is truly appropriate we shall
leave as a matter for future investigation (i.e., finding such a
quantity for conjecture~2 and proving it, or somehow show that there
is no such notion for which conjecture~2 will hold).

Analyzing the dust-ball spacetimes ($k_{\text{RW}}=+1$,
$k_{\text{RW}}=0$, and $k_{\text{RW}}=-1$) we find that conjecture~1
always holds with $k \ge 4 + \pi$.  (Again, this lower bound on $k$ is
not least and it is likely that the conjecture holds for all $k \ge
2\pi$.)  So, from these observations, it would seem that
$M_{\text{ADM}}$ or the future asymptotic values of $M_{\text{irr}}$
or $m$ could be appropriate for the mass in conjecture~1.  It is in
the proof of conjecture~1 in the spherically symmetric case (i.e.,
theorem~1) that we discover that $M_{\text{irr}}$ is the most natural
candidate, which is why this is the quantity chosen for conjecture~1.

\section{Generalizing the topology in conjecture~2} \label{sec:topology}
In this section we investigate the problem of relaxing the
restriction to ${\Bbb R}^3$ Cauchy surfaces in Conjecture~2.

As we are interested in asymptotically flat spacetimes, we begin by
restricting ourselves to manifolds where all of the ``interesting
topology'' is bounded away from infinity.  We make this idea precise
by demanding that the Cauchy surfaces have the topology of a
connected, orientable, three-manifold without boundary $C$ less a
point which we label $i^0$.  The idea here is that deleted point $i^0$
``represents'' spatial infinity.  So, outside of a sufficiently large
sphere (i.e., inside of a two-sphere sufficiently near $i^0$) the
space has the topology ${\Bbb R} \times S^2$.  Further, to exclude the
possibility of escaping to other asymptotic regions, as in the
maximally extended Schwarzschild spacetime, we demand that $C$ be
compact.  This way all of the non-compactness of a Cauchy surface
$\Sigma$ is due to the single asymptotic region.  We summarize these
requirements by demanding that $\Sigma \approx C\setminus i^0$, where
$C$ is a connected, orientable, closed (compact without boundary)
three-manifold.  As a simple example, $\Sigma \approx {\Bbb R}^3$ is
of this form with $C \approx S^3$.

Restricting ourselves to spacetimes with Cauchy surfaces diffeomorphic
to $C\setminus i^0$, does conjecture~2 hold for all closed
three-manifolds $C$?  Remarkably, the answer is no.  We can construct
a spacetime meeting the conditions of Conjecture~2 with $C = T^3 = S^1
\times S^1 \times S^1$ (the three-torus) and yet possessing infinitely
long timelike curves within the black hole as follows \cite{Brill83}.
In constructing the $k=0$ dust-ball spacetimes, we took the region
lying inside of a spherically symmetric timelike three-surface in a
$k=0$ Robertson-Walker spacetime and attached it to Schwarzschild as
we wanted the model to have ${\Bbb R}^3$ Cauchy surfaces.  Taking the
region lying outside of this timelike three-surface and attaching it
to Schwarzschild, we obtain a globally hyperbolic asymptotically flat
spherically symmetric spacetime with ${\Bbb R} \times S^2$ Cauchy
surfaces and infinitely long lived observers within the black
hole---just like in Schwarzschild.  The spatial topology still has the
form $C\setminus i^0$, although $C$ here being ${\Bbb R}^3$ is not
compact.  However, the spatial homogeneity of the Robertson-Walker
spacetimes allows an alternative construction that gives us black-hole
spacetimes with $C$ closed.  From a $k=0$ Robertson-Walker spacetime
we construct a new spacetime by simply identifying the the spacetime
under a discrete group of isometries so that the spatial topology is
now $T^3$.  Now choosing a (locally spherically symmetric) two-sphere
in this spacetime and generating a timelike three-surface by carrying
the sphere along the flow of the dust and then attaching the exterior
of this three-surface appropriately to a portion of maximally extended
Schwarzschild, we again have a globally hyperbolic asymptotically flat
spacetime, but now the Cauchy surfaces are diffeomorphic to $T^3$
minus a point. This spacetime admits infinitely long timelike curves
in the black-hole region and therefore, independent of what is chosen
for $M_\Sigma$ (as long as it finite) there is no finite $k$ for which
conjecture~2 would hold for Cauchy surfaces diffeomorphic to $T^3
\setminus i^0$.  (Note that we can perform a similar construction on
the $k=-1$ Robertson-Walker spacetimes.)

So, as conjecture~2 does not hold for all Cauchy surfaces of the form
$C\setminus i^0$ with $C$ closed, are there any restrictions on the
allowed topologies of $C$ that will give a generalized Conjecture~2 a
chance to hold?  A hint is provided by the realization that in the
study of the closed-universe recollapse conjecture, a restriction on
the Cauchy surface topology is necessary as otherwise there may be a
topological obstruction to the existence of a maximal Cauchy surface.
Furthermore, in the case where a maximal Cauchy surface exists, there
is a finite upper bound to the lengths of all causal curves in the
spacetime (provided certain energy and genericity conditions are met)
\cite{MarsdenTipler80}.  Does a similar result hold for the
asymptotically flat spacetimes?  That is, if the spacetime does admit
a maximal Cauchy surface, then must there be an upper bound to the
lengths of timelike curves within the black-hole region?  Whether this
is the case is unknown, however, adapting the ideas of Bartnik
\cite{Bartnik88,Bartnik89}, we can bound the lengths of timelike
curves in the portion of the black hole that is not reachable from the
asymptotic region.

Consider the portion of the black hole that is so deep within that it
is ``hidden'' in the sense that it cannot be reached from
$\langle\langle{\cal J}\rangle\rangle$, i.e.,
\begin{equation} \label{hidden}
M \setminus (J^-({\cal J}^+) \cup J^+({\cal J}^-)).
\end{equation}
This region is globally hyperbolic (provided the entire spacetime is
globally hyperbolic) and is simply the intersection of the black-hole
and white-hole regions associated with an asymptotic end.  In
Schwarzschild, this is all of region IV in Fig.~\ref{schwarz1}.
In the dust-ball spacetime of Fig.~\ref{db1}, this set is empty.
In the dust-ball spacetime of Fig.~\ref{db2}, this is the union of
regions $\text{I}'''$, $\text{II}'$, $\text{III}'$, and IV.  Note that
this is precisely the region of the black hole that is not covered by
conjecture~1.  Furthermore, for the spacetimes with Cauchy surfaces
diffeomorphic to a closed three-manifold minus a point, these regions
are very much like closed universes in that for any Cauchy surface
$\Sigma$, the domain of dependence of the compact set $\Sigma
\setminus \langle\langle{\cal J}\rangle\rangle$ contains the hidden
black-hole region.  Therefore, the evolution of these regions is
determined by data on a compact set.  This property provides another
connection between the conjectures that observers in closed-universes
and black-hole regions must have finite lifetimes and using it we now
show that if the spacetime admits a maximal Cauchy surface, then the
lengths of all causal curves in this hidden portion of the black hole
are bounded above.

{\it Theorem 3.}  Fix an asymptotically flat spacetime satisfying the
timelike-convergence condition \cite{ec} and possessing a maximal
Cauchy surface $\Sigma_0$ diffeomorphic to $C\setminus i^0$ where $C$
is a connected, orientable, closed three-manifold.  Then, there exists
an upper bound on the lengths of all causal curves in the hidden
portion of the black hole (i.e., the region given by
Eq.~(\ref{hidden})).

{\it Proof.}  Denoting the hidden portion of the black hole by $H$,
our task is to show that $d(H,H)$ is finite.  Since $d(H,H) \le
d(H,\Sigma_0) + d(\Sigma_0,H)$, we need only show that $d(H,\Sigma_0)$
and $d(\Sigma_0,H)$ are finite.  [Recall that $d(p,q)$ is the distance
function defined as the least upper bound to the length of all
continuous causal curves connecting $p$ to $q$ if $q \in J^+(p)$, and
zero otherwise, and $d({\cal P},{\cal Q})$ is defined for subsets
${\cal P}$ and ${\cal Q}$ of $M$ as the least upper bound of $d(p,q)$
over all $p \in {\cal P}$ and $q \in {\cal Q}$ \cite{HawkingEllis73}.]

We bound $d(\Sigma_0,H)$ as follows.  Fix any point $p \in
D^+(\Sigma_0) \cap H$ and let $\gamma$ be a longest timelike curve
from $\Sigma_0$ to $p$.  As is well known, $\gamma$ intersects
$\Sigma_0$ orthogonally, is geodetic, and has no points between
$\Sigma_0$ and $p$ conjugate to $\Sigma_0$.  (See, for example,
Sec.~9.3 of Ref.~\cite{Wald84}.)\ \ However, this last requirement is
difficult to satisfy for long curves as a congruence of timelike
geodesics meeting $\Sigma_0$ orthogonally has vanishing expansion on
$\Sigma_0$ since $\Sigma_0$ has zero trace extrinsic curvature $K$ by
virtue of its being a maximal hypersurface.  Were $K$ everywhere
bounded away from zero by a negative constant $-\kappa$ on $\Sigma_0$,
or merely on $J^-(H) \cap \Sigma_0$, then the length of $\gamma$
(which equals $d(\Sigma_0,p)$) could be no larger than $3/\kappa$.
(See theorem~9.5.1 of Ref.~\cite{Wald84}.)\ \ However, $K$ is merely
zero on $\Sigma_0$.  To take care of this problem, we construct a
Cauchy surface $\Sigma_1$ to the future of $\Sigma_0$ having negative
trace extrinsic curvature bounded away from zero on $J^-(H) \cap
\Sigma_1$ by ``pushing'' or ``evolving'' $\Sigma_0$ to the future.

We construct $\Sigma_1$ by first constructing a family of Cauchy
surfaces $\Sigma_t$ by evolving $\Sigma_0$ with zero shift (so the
displacement is purely normal) and non-constant lapse $N$.  The trace
of the extrinsic curvature of the surface $\Sigma_t$ then evolves
according to
\begin{equation} \label{dKdt}
{{\partial K} \over {\partial t}} = - N (K_{ab}K^{ab} + R_{ab}n^a n^b) +
D_a D^a N,
\end{equation}
where $n_a = -N(dt)_a$ is the future-directed unit normal to the
surfaces, $K_{ab}$ is the extrinsic curvature of the surfaces, and
$D_a$ is the derivative operator associated with the induced metric on
each surface.  Note that with $N=1$, the right-hand side of
Eq.~(\ref{dKdt}) is always non-negative as $K_{ab}K^{ab}$ is
manifestly non-negative and $R_{ab}n^a n^b$ is non-negative by the
timelike-convergence condition.  Were the sum of these quantities
positive everywhere on $\Sigma_0$, then by simply evolving $\Sigma_0$
with $N=1$, we would have a Cauchy surface with $K < 0$ for any $t >
0$.  However, this need not be the case, so we need to be a bit more
subtle in our choice of $N$.

Let $Y_0$ be any compact subset of $\Sigma_0$ (with smooth boundary)
whose interior contains the compact set $X_0 = \Sigma_0\setminus
J^+({\cal J}^-)$.  Then, $(J^-(p) \cap \Sigma_0) \subset X_0 \subset
Y_0$.  There is a great deal of freedom in our construction of
$\Sigma_t$ and hence $\Sigma_1$, but for definiteness, we shall
consider any family of hypersurfaces $\Sigma_t$ that arises from
$\Sigma_0$ and a lapse that obeys the property that, on $\Sigma_0$,
$N$ is the unique solution of the elliptic partial differential
equation $D_a D^a N = -f$ for some scalar field $f$ that is positive
in the interior of $Y_0$ and zero on $\Sigma_0\setminus Y_0$ and that
satisfies the boundary condition that $N$ be zero on
$\Sigma_0\setminus Y_0$.  By Hopf's maximum principle
\cite{ProtterWeinberger84}, $N$ is positive everywhere in the
interior of $Y_0$, so by Eq.~(\ref{dKdt}), ${\partial K}/{\partial t}
\le -f$ on $Y_0$ and therefore ${\partial K}/{\partial t} \le -c$
on $X_0$ for some positive constant $c$.  So, for sufficiently small
$t$, on $\Sigma_t\setminus J^+({\cal J}^-)$ the trace of the extrinsic
curvature $K$ is bounded away from zero by the negative constant
$-\kappa = -ct$.  We shall take $\Sigma_1$ to be any such surface.
So, as $J^-(p) \cap \Sigma_1$ is a subset of of $\Sigma_1\setminus
J^+({\cal J}^-)$ and $K \le -\kappa$ thereon, by the argument above,
$d(\Sigma_1,p) \le 3/\kappa$ and therefore $d(\Sigma_1,H) \le
3/\kappa$.  Noting that $d(\Sigma_0,H) \le d(\Sigma_0,\Sigma_1) +
d(\Sigma_1,H)$ and the fact that $d(\Sigma_0,\Sigma_1)$ is finite (as
$\Sigma_0$ and $\Sigma_1$ coincide outside of compact sets) we have
established the existence of an upper bound on $d(\Sigma_0,H)$.

By a time-reversed argument, a similar bound can be established on
$d(H,\Sigma_0)$.  Therefore, we have established the existence of a
finite upper bound on $d(H,H)$ as was to be shown.~$\Box$

So, should the spacetime admit a maximal Cauchy surface, then there
will not be arbitrarily long timelike curves contained within the
hidden portion of the black hole.  While sufficient conditions have
been given for the existence of such surfaces \cite{Bartnik84,BCM90},
not all asymptotically flat spacetimes admit maximal Cauchy surfaces.
Indeed, the Cauchy surface topology may provide an obstruction to the
existence of such a surface.  This follows from the simple fact that
scalar constraint equation of general relativity demands that the
Ricci scalar curvature associated with the induced metric on such a
surface must be non-negative provided the spacetime satisfies the
non-negative-energy condition \cite{ec} while most three-manifolds of
the form $C\setminus i^0$ do not admit Riemannian metrics with
non-negative scalar curvature \cite{Witt86}.  For instance, our
$T^3\setminus i^0$ black-hole spacetime above does not admit a maximal
Cauchy surface because of this obstruction.

What three-manifolds $C \setminus i^0$ (with $C$ as before) admit
asymptotically flat metrics of non-negative scalar curvature, i.e.,
those that do not provide a topological obstruct the existence of a
maximal Cauchy surface?  Witt has shown that any such manifold $C$
also admits a Riemannian metric with positive scalar curvature
\cite{Witt95}.  It then follows from the work of Gromov and Lawson
that $C$ must be $S^3$, $S^1 \times S^2$, or a manifold that can be
constructed from one of these by making certain identifications and
connected summations \cite{GromovLawson83}.  Therefore, we generalize
the Cauchy surface topology in conjecture~2 to include each of these
manifolds less a point.

While theorem~3 does support the plausibility of conjecture~2 (in that
some bound should exist), it would be nice to strengthen its results
beyond the hidden portion of the black hole.  However, with the
current method of proof, it does not appear that such a generalization
is possible.  The problem is that $I^-(B) \cap \Sigma_0$ need not be
contained in a compact subset of $\Sigma_0$ and therefore the
construction of a Cauchy surface $\Sigma_1$ that is contracting at a
rate bounded away from zero on subset that acts as a Cauchy surface
for the region of the black hole lying to the future of $\Sigma_1$
will fail.  The dust-ball spacetimes in Figs.~\ref{db1} and \ref{db2}
provides simple examples of this behavior.  However, in these
spacetimes, for a timelike curve in the black-hole region, the closure
of $I^-(\gamma) \cap \Sigma_0$ is a compact subset of $\Sigma_0$.  So,
here we could show that there are no infinite length causal curve
(although we would not have an upper bound independent of the curve).
Can this weaker result be shown to hold?  No.  There is no (known)
reason why $I^-(\gamma)$ must meet a Cauchy surface within a compact
subset thereof as happens when $\gamma$ ``runs into'' a Cauchy horizon
(as seen in Kerr-Newman).  Lastly, we note that the lengths of causal
curves in the region of the black hole $B$ lying to the past of
$\Sigma_0$ are bounded as $J^+(B) \cap \Sigma_0$ is the compact set $B
\cap \Sigma_0$.  So, a timelike curve in a black-hole region of a
spacetime satisfying the timelike-convergence condition and admitting
a maximal Cauchy surface cannot have infinite length to the past.

\section{The spherically symmetric case} \label{sec:spherical}
Our strategy in proving theorems~1~and~2 is quite simple.  The idea is
to first establish a number of properties of $r$ and $m$ on the event
horizon.  Next, these properties are used to establish an upper bound
for $r$ in the black-hole region (or just in the part lying to the
future of ${\cal J^-}$).  Once this is done, the theorem follows quite
easily with the aid of the following lemma established earlier in the
analysis of the closed-universe recollapse conjecture (and whose proof
is essentially the subject of Sec.~III of Ref.~\cite{Burnett95}).

{\it Lemma 1.}  In a globally hyperbolic spherically symmetric
spacetime satisfying the non-negative-pressures condition \cite{ec}
and on which $m$ is everywhere non-negative and $r$ is everywhere
bounded above by some constant $r_U$, the lengths of all timelike
curves are bounded above by $10 r_U$.

In Sec.~\ref{sec:elementaries}, the elementary features of the
spherically symmetric spacetime are reviewed.  In
Sec.~\ref{sec:r_and_m}, some properties of $r$ and $m$ associated with
asymptotic flatness are established.  In Sec.~\ref{sec:bounding_r_B},
bounds for $r$ on the black-hole region are established in a variety
of circumstances.  Lastly, in Sec.~\ref{sec:complete}, the proofs of
theorem~1 and~2 are completed.

\subsection{Elementaries} \label{sec:elementaries}

In this section, the elementary features of the spherically symmetric
spacetimes needed here are briefly reviewed.  For a more complete
presentation, see Ref.~\cite{Burnett93}.

A spacetime $(M,g_{ab})$ is said to be spherically symmetric if it
admits a group $G \approx \text{SO}(3)$ of isometries, acting
effectively on $M$, each of whose orbits is either a two-sphere or a
point.  Denote the orbit of a point $p$ by ${\cal S}_p$.  The value of
the non-negative scalar field $r$ at each $p \in M$ is defined so that
$4\pi r^2$ is the area of ${\cal S}_p$.  So, in particular, $r(p)=0$
if ${\cal S}_p = p$, while $r(p)>0$ if ${\cal S}_p$ is a two-sphere.
Furthermore, we shall say that ${\cal S}$ is a sphere of symmetry if
${\cal S} = {\cal S}_p$ for some $p \in M$ and ${\cal S}$ is a
two-sphere.

Where $r>0$, we decompose the metric $g_{ab}$ into the sum $g_{ab} =
h_{ab} + q_{ab}$, where $q^a{}_b$ is the projection operator onto the
tangent space of each sphere of symmetry and $h^a{}_b$ is the
projection operator onto the tangent space of each two-surface
perpendicular to the spheres of symmetry.  Using the fact that there
exists a preferred ``unit-metric'' $\Omega^{ab}$ on each sphere of
symmetry, we have $q_{ab} = r^2 \Omega_{ab}$ (where
$\Omega^{am}\Omega_{mb} = q^a{}_b$ and $\Omega_{ab} = q^m{}_a q^n{}_b
\Omega_{mn}$).  This gives us the final decomposition of $g_{ab}$ as
\begin{equation}
g_{ab} = h_{ab} + r^2 \Omega_{ab}.
\end{equation}

For the spherically symmetric spacetimes, the mass $m$ is defined by
\begin{equation} \label{defm}
2m = r(1-\nabla_m r \nabla^m r).
\end{equation}
Defining $\epsilon^{ab}$ to be either of the two antisymmetric
tensor fields such that $\epsilon^{ab}\epsilon^{cd} = -2
h^{a[c}h^{d]b}$, and denoting the ``radial part'' of the Einstein
tensor $G^{ab}$ by $\tau^{ab}$ (i.e., $\tau^{ab}=h^a{}_m h^b{}_n
G^{mn}$) we have
\begin{equation} \label{DDr}
h_a{}^m h_b{}^n \nabla_m \nabla_n r = {m \over r^2} h_{ab} - {r \over 2}
\tau^{mn}\epsilon_{ma}\epsilon_{nb},
\end{equation}
from which it follows that
\begin{equation} \label{D2m}
\nabla_a (2m) = r^2 \tau^{mn} \epsilon_{ma}\epsilon_{nb} \nabla^b r.
\end{equation}

As the space of vectors perpendicular to the spheres of symmetry is
two-dimensional and the metric thereon is Lorentzian, we shall label a
radial vector as being either outgoing, ingoing, future-directed, or
past-directed according to which quadrant the vector lies with the
convention that $\nabla^a r$ is outgoing near ${\cal J}$.  Radial null
vectors will have two such labels (e.g., outgoing and future-directed)
while the zero vector will have all four.  (In other words, the set of
vectors under a particular label is closed.)\ \ Note that these
designations are well defined globally for any spherically symmetric
spacetime that is simply connected as are the spherically symmetric
spacetimes with ${\Bbb R}^3$ or ${\Bbb R} \times S^2$ Cauchy surfaces.

\subsection{Properties of $r$ and $m$ associated with asymptotic flatness}
\label{sec:r_and_m}

Our first lemma establishes a few basic properties of $r$.

{\it Lemma 2.}  Fix a globally hyperbolic asymptotically flat
spherically symmetric spacetime satisfying the null-convergence
condition \cite{ec}.  Then: (a) on $J^-({\cal J}^+)$, $\nabla^a r$
cannot be future-directed or ingoing; (b) on $J^+({\cal J}^-)$,
$\nabla^a r$ cannot be past-directed or ingoing; (c) on
$\langle\langle{\cal J}\rangle\rangle$, $\nabla^a r$ must be outgoing
spacelike; (d) on $\partial J^-({\cal J}^+)$, $\nabla^a r$ must be
either past-directed or outgoing and $r$ is non-decreasing to the
future.

{\it Proof.} To prove (a), fix any sphere of symmetry ${\cal S}
\subset J^-({\cal J}^+)$.  If $\nabla^a r$ where future-directed or
ingoing on ${\cal S}$, then ${\cal S}$ would be a future outer trapped
surface in $J^-({\cal J}^+)$ which is impossible by Proposition~9.2.8
of Ref.~\cite{HawkingEllis73}.  Part (b) follows from (a) by simply
reversing the time-orientation of the spacetime.  Part (c) follows
from (a) and (b) and noting that outgoing spacelike is the only
possibility left.  That $\nabla^a r$ must be either past-directed or
outgoing in (d) follows from part (a) and continuity from which it
follows that $r$ is non-decreasing to the future.  (Alternatively,
part (d) follows from Hawking's black-hole area theorem.)~$\Box$

Our next lemma establishes properties of $m$ in the asymptotic region
and on the event horizon.

{\it Lemma 3.} Fix a globally hyperbolic asymptotically flat
spherically symmetric spacetime satisfying the dominant-energy
condition \cite{ec}. Then, on $\langle\langle{\cal J}\rangle\rangle$,
$m \le M_{\text{ADM}}$.  Further, on the portion of the event horizon
lying the future of ${\cal J^-}$ [i.e., $(\partial J^-({\cal J}^+))
\cap J^+({\cal J^-})$], $m$ is non-decreasing to the future and has a
future limit which is no greater than $M_{\text{ADM}}$.

{\it Proof.}  As $\nabla^a r$ is outgoing spacelike on
$\langle\langle{\cal J}\rangle\rangle$ by lemma~2, $\nabla^a m$ is
outgoing thereon by Eq.~(\ref{D2m}).  Fix any point $p \in
\langle\langle{\cal J}\rangle\rangle$, and a spherically symmetric
Cauchy surface $\Sigma$ containing $p$.  As $m$ is non-decreasing
outwards on $\Sigma \cap \langle\langle{\cal J}\rangle\rangle$ and
since $m \rightarrow M_{\text{ADM}}$ as we approach $i^0$ in $\Sigma$,
$m \le M_{\text{ADM}}$ at $p$ and hence on all of $\langle\langle{\cal
J}\rangle\rangle$.  By lemma~2, $\nabla^a r$ must be outgoing (and yet
not past-directed) on $(\partial J^-({\cal J}^+)) \cap J^+({\cal
J^-})$, and hence $\nabla^a m$ must be outgoing thereon.  Therefore,
$m$ must be non-decreasing to the future thereon.  By continuity, $m$
is bounded by $M_{\text{ADM}}$ on $(\partial J^-({\cal J^+})) \cap
J^+({\cal J^-})$ and so $m$ must have a limit no greater than
$M_{\text{ADM}}$ to the future.~$\Box$

It is worth noting that the upper bound of $M_{\text{ADM}}$ on $m$ on
the event horizon need not hold everywhere on the event horizon.  A
simple example is provided by the dust-ball spacetime in
Fig.~\ref{db2}.  However, on the portion of the event horizon lying to
the future of some Cauchy surface, $m$ is bounded as follows.

{\it Lemma 4.}  Fix a globally hyperbolic asymptotically flat
spherically symmetric spacetime satisfying the non-negative-pressures
condition \cite{ec}.  Fix a Cauchy surface $\Sigma$ therein and denote
by $P$ the subset of $\Sigma$ on which $\nabla^a r$ is past-directed
timelike, past-directed null, or zero.  On the portion of the
black-hole's event horizon ${\cal E}$ in the future domain of
dependence of $\Sigma$, i.e., $D^+(\Sigma) \cap {\cal E}$, $m$ is
bounded above by the expression
\begin{equation} \label{lemma_4}
2m \le \max \biglb( \sup_{{\cal E}}(r), \sup_P(2m) \bigrb).
\end{equation}

{\it Proof.}  By lemma~2, $\nabla^a r$ is either outgoing or
past-directed on ${\cal E}$.  Where $\nabla^a r$ is outgoing, then $2m
\le r \le \sup_{\cal E}(r)$, where the first inequality follows
directly from the definition of $m$.  Where $\nabla^a r$ is
past-directed, by lemma~C1 of Appendix~C, $2m \le \sup_P(2m)$.
Combining these results, Eq.~(\ref{lemma_4}) follows.~$\Box$

Lastly, that $m$ is everywhere non-negative for the spacetimes under
consideration is given by the following lemma.

{\it Lemma 5.} On a spherically symmetric spacetime satisfying the
dominant-energy condition \cite{ec} and admitting a complete
spherically symmetric Cauchy surface $\Sigma$, $m$ is everywhere
non-negative.

{\it Proof.}  We first show that $m$ is non-negative on $\Sigma$ and
then show that the non-negativity of $m$ everywhere else follows.
Suppose that at some point $p \in \Sigma$ that $m(p)<0$.  Then, by
Eq.~(\ref{defm}), $\nabla^a r$ is necessarily spacelike at $p$.
Denote the subset of $\Sigma$ where $r>0$ by $\tilde{\Sigma}$ (which
is just $\Sigma$ less one, two, or possibly no points) and let $s^a$
the spherically symmetric unit-vector field on $\tilde{\Sigma}$ such
that $s^a \nabla_a r < 0$ at $p$.  Consider the maximal integral curve
$\sigma$ of $s^a$ emanating from $p$.  Then, as $s^a \nabla_a (2m) \le
0$ where $\nabla^a r$ is spacelike with $s^a \nabla_a r < 0$, $m$ is
negative and $\nabla^a r$ is spacelike all along $\sigma$.  Therefore,
along $\sigma$, the change in $r$ with respect to the distance $s$
measured along $\sigma$ from $p$ is bounded by $dr/ds = s^a \nabla_a r
\le - \sqrt{\nabla^a r \nabla_a r} = -\sqrt{1-2m/r} < -1$.
Integrating this and using the fact that $r>0$ on $\tilde{\Sigma}$,
the length of $\sigma$ is bounded above by $r(p)$.  The curve $\sigma$
is geodetic (in $\Sigma$) and, furthermore, it cannot be extended to a
greater length in $\Sigma$ as $m$ is zero where $r$ is zero.
Therefore, with $\Sigma$ complete, $m$ must be non-negative everywhere
on $\Sigma$.

Having established that $m$ is non-negative on $\Sigma$, we now show
that $m$ is non-negative everywhere as follows.  Suppose that $m<0$ at
some point $p \in D^+(\Sigma)$.  Then, $\nabla^a r$ is necessarily
spacelike at $p$.  So, consider the past-directed radial null geodesic
$\nu$ with future endpoint $p$ such that $k^a \nabla_a r < 0$ at $p$,
where $k^a$ is a past-directed tangent vector to $\nu$.  Then, either:
(i) $\nabla^a r$ remains spacelike with $r>0$ on $\nu$ in which case
it must intersect $\Sigma$ at some point $q$.  So, as $k^a
\nabla_a(2m) \le 0$, we have $m(p) \ge m(q) \ge 0$; or (ii) $\nabla^a
r$ remains spacelike on $\nu$ but it intersects the ``center of
symmetry'' where $r=0$ at some point $q$, so that $m(p) \ge m(q) = 0$;
or (iii) at some point $q$ on $\nu$, $\nabla^a r$ fails to remain
spacelike in which case it must be null or zero so that $m(p) \ge m(q)
= r(q)/2 > 0$.  In all three cases, $m(p)$ is non-negative, so $m$ is
non-negative on $D^+(\Sigma)$.  That $m$ is non-negative on
$D^-(\Sigma)$ follows by a similar argument.~$\Box$

\subsection{Bounding $r$ inside the black hole} \label{sec:bounding_r_B}

Here we establish bounds on $r$ in the black-hole region under a
variety of circumstances.

{\it Lemma 6.}  Fix a globally hyperbolic asymptotically flat
spherically symmetric spacetime satisfying the null-convergence
condition \cite{ec} and possessing a black hole $B$ with event horizon
${\cal E} = \partial B$.  Fix a spherically symmetric Cauchy surface
$\Sigma$ for this spacetime and denote by $\Sigma'$ the component of
$\Sigma$ connected to $i^0$ on which $\nabla^a r$ is outgoing.  Then,
\begin{equation} \label{lemma_6}
r \le \sup_{\cal E}(r)
\end{equation}
on: (a) that portion of the black hole lying in the future of the past
null-infinity [i.e., on $B \cap J^+({\cal J^-})$]; (b) on the portion
of the black hole lying in the future of $\Sigma'$ [i.e., on $B \cap
J^+(\Sigma')$]; (c) on all of the black-hole region provided that
$\nabla^a r$ is outgoing on all of $\Sigma$.

{\it Proof.}  To prove (a), fix any point $p \in B \cap J^+({\cal
J^-})$.  Let $\nu$ be an outgoing null geodesic generator of $\partial
J^-({\cal S}_p)$.  Then $\nu$ has a future endpoint on ${\cal S}_p$
and no past endpoint.  Further, it can be shown that $\nu$ is
contained in $J^+({\cal J^-})$ and enters $\langle\langle{\cal
J}\rangle\rangle$ and hence must intersect ${\cal E}$.  Therefore, as
$\nabla^a r$ is not past-directed or ingoing on $J^+({\cal J^-})$ by
lemma~2, $r$ is decreasing to the future on $\nu$ and hence $r(p) \le
r(\nu \cap {\cal E}) \le \sup_{\cal E}(r)$.

To prove (b), fix any point $p \in B \cap J^+(\Sigma')$.  If $r(p)$ is
zero, then Eq.~(\ref{lemma_6}) is immediate.  Otherwise, consider the
(unique) radial ingoing and future-directed null geodesic $\nu$ with
past endpoint $q$ on $\Sigma'$ and future endpoint $p$.  Denote by
$\lambda$ and $l^a$ the parameter and associated tangent vector,
respectively, in an affine parameterization of $\nu$.  Then,
$dr/d\lambda = l^a \nabla_a r \le 0$ at $q$ as $\nabla^a r$ is
outgoing at $q$.  Further, along $\nu$, $dr/d\lambda$ remains negative
as can be seen as follows.  Consider,
\begin{equation} \label{Raychaudhuri}
{d^2 r \over d\lambda^2}  =  l^a \nabla_a(l^b \nabla_b r)
                          =  l^a l^b \nabla_a \nabla_b r
                         \le 0.
\end{equation}
The second equality follows from the fact that $\nu$ is geodetic.  The
inequality follows from Eq.~(\ref{defm}), the fact that $l^a$ is null,
that $\epsilon^a{}_b l^b = \pm l^a$ (sign depending on one's choice of
$\epsilon_{ab}$), and the fact that $\tau_{ab}$ satisfies the
null-convergence condition.  [Note that Eq.~(\ref{Raychaudhuri}) is
the Raychaudhuri equation for a congruence of radial null geodesics
where their expansion $\theta$ is given by $\theta = {2 \over r} { dr
\over d\lambda}$ \cite{HawkingEllis73,Wald84}.]\ \ Therefore, $r$ is
non-increasing to the future along $\nu$.

Should $\nu$ intersect the event horizon, then $r(p)$ is bounded above
by $r(\nu \cap {\cal E})$ which in turn is bounded above by
$\sup_{\cal E}(r)$.  Otherwise, $\nu$ has its past endpoint on
$\Sigma' \cap B$, so as $\nabla^a r$ is outgoing on $\Sigma'$, $r$
on $\Sigma' \cap B$ is bounded above by $r$ on its outer boundary,
which gives $r(p) \le \max_{\Sigma' \cap B}(r) = r({\cal E} \cap
\Sigma) \le \sup_{\cal E}(r)$.

Finally, to prove (c), we note that $\Sigma' = \Sigma$ and so by part
(b), Eq.~(\ref{lemma_6}) holds on $B \cap J^+(\Sigma)$.  To establish
this bound on $B \cap J^-(\Sigma)$, fix any point $p$ therein.  The
proof proceeds identically to the above except now $\nu$ is a
past-directed ingoing radial null geodesic with past endpoint $p$ and
future endpoint $q$ on $\Sigma$.  [Note that $q \in B$ since $p \in B$
and $q \in J^+(p)$.]\ \ This time $r$ is non-increasing to the past
along $\nu$ giving us the bounds $r(p) \le r(q) \le \sup_{B \cap
\Sigma}(r) = r({\cal E} \cap \Sigma) \le \sup_{\cal E}(r)$.~$\Box$

{\it Lemma 7.}  Fix an asymptotically flat spherically symmetric
spacetime that possesses ${\Bbb R}^3$ Cauchy surfaces and that
satisfies the dominant-energy and non-negative-pressures condition
\cite{ec}.  Then for any spherically symmetric Cauchy surface $\Sigma$
therein, on the black-hole region $B$ with event horizon ${\cal E} =
\partial B$, we have
\begin{equation} \label{bound_r_3}
r \le \max \biglb( { \sup_{\cal E}(r),
\max_{(\Sigma \cap B) \cup P}(2m) } \bigrb),
\end{equation}
where $P$ is the subset of $\Sigma$ on which $\nabla^a r$ is
past-directed timelike, past-directed null, or zero.

{\it Proof.}  We first establish Eq.~(\ref{bound_r_3}) for the portion
of the black hole lying on or to the future of $\Sigma$.  For this
region, the surface $\Upsilon$ defined as the union of $\Sigma \cap B$
together with $D^+(\Sigma) \cap {\cal E}$, acts as a Cauchy surface
for $D^+(\Sigma) \cap B$ as any past-directed causal curve with future
endpoint $p \in D^+(\Sigma) \cap B$ and no past endpoint must
intersect $\Upsilon$.  To see this, for any such point $p$, consider
any past-directed causal curve $\lambda$ with future endpoint $p$ and
without past endpoint.  The curve $\lambda$ must intersect $\Sigma$ at
some point $q$ as $p \in D^+(\Sigma)$.  If $q \in B$, then $q \in
\Sigma \cap B$.  If $q \notin B$, then there must a point on $\lambda$
between $p$ and $q$ that lies on ${\cal E}$ and thus $\lambda$
intersects $D^+(\Sigma) \cap {\cal E}$.  Therefore, appealing to
lemma~3 of Ref.~\cite{Burnett93} (or more correctly just repeating the
argument used therein as our region is not quite a spacetime in that
it has boundaries and further noting that the dominant-energy
requirement was superfluous), on $B$ we have
\begin{equation}
r \le \max \biglb( \sup_\Upsilon (r), \sup_\Upsilon (2m) \bigrb).
\end{equation}
However, we have
\begin{mathletters}
\begin{eqnarray}
\max_{\Sigma \cap B}(r) & \le & \max \biglb( \sup_{\cal E}(r),
\max_{\Sigma \cap B} (2m) \bigrb), \label{b2} \\
\sup_{D^+(\Sigma) \cap {\cal E}} (2m) & \le &
\max \biglb( \sup_{\cal E}(r),
\max_{P} (2m) \bigrb). \label{b3}
\end{eqnarray}
\end{mathletters}
Equation~(\ref{b2}) follows from lemma~C2 in Appendix~C by taking $C =
\Sigma \cap B$ and noting that the boundary of $C$ is a subset of
${\cal E}$.  Equation~(\ref{b3}) follows from lemma~4.  Combining
these results, Eq.~(\ref{bound_r_3}) follows.  (Note: It is in the
establishment of Eq.~(\ref{b2}) that our restriction to ${\Bbb R}^3$
Cauchy surfaces is used.  Had ${\Bbb R} \times S^2$ Cauchy surfaces
been allowed, $C$ would not have been compact and $r$ may be unbounded
on $\Sigma \cap B$ as it is in the maximally extended Schwarzschild
spacetime.)

We now establish Eq.~(\ref{bound_r_3}) for the portion of the black
hole lying on or to the past of $\Sigma$.  For this region, the
surface $\Sigma \cap B$ acts as a Cauchy surface for $D^-(\Sigma) \cap
B$ in the sense that any future-directed causal curve with past
endpoint $p \in D^-(\Sigma) \cap B$ and no future endpoint must
intersect $\Sigma \cap B$.  (Any such curve $\lambda$ must intersect
$\Sigma$ at some point $q$ as $p \in D^-(\Sigma)$.  However, $q \in B$
as $p \in B$ and hence $\lambda$ intersects $\Sigma \cap B$.)\ \ The
proof now follows as before with $\Upsilon = \Sigma \cap B$.~$\Box$

\subsection{Completing the proofs of theorems 1 and 2} \label{sec:complete}

To complete the proofs of theorem~1~and~2, we simply observe that
theorem~1 follows from lemmas~1, 5, and~6, while theorem~2 follows
from lemmas~1, 5, and~7.

\section{What can be done?} \label{sec:do}
A massive spherical shell surrounds a spherical vacuum (Schwarzschild)
black hole so that $m=M_1$ inside of the shell and $m=M_2 \gg M_1$
outside of the shell.  An observer begins his journey on a future
marginally outer trapped sphere of symmetry ${\cal S}$ (which will be
on the event horizon if the shell of matter does not fall into the
black hole).  If the shell does not collapse becoming trapped within
the black-hole region, then theorem~1 [part (c)] guarantees that the
lifetime of our traveler is bounded above by $20 M_1$, which is on the
order of the true maximum lifetime $\pi M_1$.  However, suppose that
the shell does fall inward creating a black hole of final mass $M_2$.
In this case, theorem~1 bounds the lifetime of the observer by $20
M_2$, which is much larger than the vacuum maximum of $\pi M_1$.  Can
the observer truly live this long?

More generally, does the fact that the bound in theorem~1 is a
multiple of the future asymptotic size of the black hole indicate that
one can increase the lifetime of an observer therein by throwing
additional matter into the hole to make it larger?  Further, as it
appears that the only natural upper bound for the size of the black
hole $M_{\text{irr}}$ is the mass of the spacetime $M_{\text{ADM}}$
(lemma~A1), which corresponds to throwing all the available matter
into the black hole (not that this is always possible), might one be
able to increase the lifetime of our traveler within the black hole to
be on the order of $M_{\text{ADM}}$?  That this is not possible in our
scenario is a consequence of the following theorem.

{\it Theorem 4.}  Fix a globally hyperbolic spherically symmetric
spacetime satisfying the dominant-energy and non-negative-pressures
conditions \cite{ec} and possessing a spherically symmetric Cauchy
surface that is geodesically complete.  Then, for any sphere of
symmetry ${\cal S}$ that is future trapped, the lengths of all causal
curves in $J^+({\cal S})$ are bounded above by $10 r({\cal S}) \le 20
m({\cal S})$.

{\it Proof.}  This theorem follows from lemma~1 and the fact that $r
\le r({\cal S})$ on $J^+({\cal S})$, which we now establish.
Fix any point $p \in I^+({\cal S})$ with $r(p)>0$ and fix any maximal
length timelike curve $\gamma$ from ${\cal S}$ to $p$.  The curve
$\gamma$ is radial and geodetic with $r$ positive thereon.  It follows
from the non-negative-pressures condition that $d^2r/dt^2 \le -m/r^2$
on $\gamma$ (where $t$ is the proper time along $\gamma$), and since
$m$ is non-negative by lemma~5, $d^2r/dt^2 \le 0$ on $\gamma$.
However, as $\nabla^a r$ is future-directed timelike, future-directed
null, or zero on ${\cal S}$, $dr/dt$ must be nonpositive on ${\cal
S}$.  Therefore, $dr/dt \le 0$ on all of $\gamma$.  Thus, $r(p) \le
r({\cal S})$.  That $r \le r({\cal S})$ on the boundary of $J^+({\cal
S})$ follows by continuity of $r$ (or by a slight modification of the
above argument for the null case).~$\Box$

This theorem bounds the length of the journey of the observer in the
scenario above by $10 r({\cal S}) \le 20 m({\cal S}) = 20 M_1$, which
shows that the additional matter cannot increase his lifetime to be on
the order of $M_2$.  In fact, it turns out that in this case (and
those like it) the additional matter never increases the lifetime of
an observer within.  (See theorem~6 below.)

To address the more general problem of whether we can increase the
lifetime of an observer within a black hole, some care is needed as
the problem is subtle for at least three reasons.  First, while it may
be possible to increase the amount of time an observer can possibly
exist within a black hole, it may just be that none of this increase
is available to the observer that is {\it already} within the
black-hole region.  That is, while we expect that observers can exist
longer in larger black holes (as is certainly true for vacuum
Schwarzschild black holes), and it is true that by throwing in matter
we can increase the amount of time an observer has spent within a
black hole (by making the black hole larger, which is related to our
next point), it may be that an observer still cannot live any longer
than he would have otherwise.  Second, the very notion of a black hole
is a nonlocal concept requiring a knowledge of the entire future
history for its determination.  For example, while the spacetime
corresponding to a static spherically symmetric shell of matter with a
flat interior has no black-hole region, if we choose to ``turn off''
the pressure that keeps the shell from collapsing inwards, the new
spacetime will possess a nonempty black-hole region.  So, while under
one evolution an observer may begin his journey outside of the
black-hole region, in another he may already be within this region.
Third, the manipulation of matter is a delicate issue in general
relativity.  We cannot move matter around as we like, e.g., pulling
two stars together or pushing them apart.  The pushing and pulling
involve stresses that must be taken into account as demanded by
Einstein's theory of gravity.

To take care of these subtleties, we formulate our problem as an
initial-value problem as follows.  Recall that an initial data set $I$
for general relativity consists of a triple $I=(\Sigma,q_{ab},K_{ab})$
\cite{Wald84}.  Here $\Sigma$ is a three-manifold (physically the
spacetime at an ``instant''), $q_{ab}$ is a Riemannian metric on
$\Sigma$ (physically the metric induced from the spacetime metric
$g_{ab}$), and $K_{ab}$ is a symmetric tensor on $\Sigma$ (physically
the extrinsic curvature of $\Sigma$ in $M$).  On $\Sigma$, $q_{ab}$
and $K_{ab}$ are not freely specifiable.  Instead, the initial-value
constraint equations of general relativity impose a relation between
these quantities and the energy density and mass-current density of
the matter fields.  (See Eqs.~(10.2.41) and~(10.2.42) of
Ref.~\cite{Wald84}.)

Fix an initial data set $I$ and a point $p$ in the three-manifold
$\Sigma$ associated with $I$.  Our knowledge of $I$ corresponds to a
knowledge of the state of the Universe at an ``instant of time'',
while our knowledge of $p$ corresponds to our knowledge of the
position of our observer at that time.  If we had a detailed
description of the matter fields (in particular their initial data
satisfying the constraint equations and the equations governing their
evolution) then with Einstein's equation we could evolve this initial
data to produce a unique spacetime (up to diffeomorphisms).  However,
we wish investigate the behavior of gravitational fields without this
detailed knowledge of the matter fields, One way around this problem
is to restrict ourselves to vacuum spacetimes.  Instead, we choose to
proceed as follows.

To restrict ourselves to spacetimes that are ``physical'' in the sense
that their matter content could be considered in some sense
``ordinary'', we shall restrict ourselves to spacetimes whose Einstein
tensor satisfies a fixed set of inequalities (energy conditions).  For
definiteness, here we shall demand that the spacetime satisfy the
dominant-energy and non-negative-pressures conditions \cite{ec}.  We
shall call a spacetime $(M,g_{ab})$ satisfying these energy conditions
a {\it possible evolution} of an initial data set $I$ if there is a
Cauchy surface $S$ in $M$ such that the induced data on $S$ coincides
with that of $I$.  Note that without the energy conditions, there will
be many possible evolutions of any given initial data set.  In fact,
we can ``evolve'' any initial data set $I_1$ to any other initial data
set $I_2$ in the sense that there will be a single spacetime that is a
possible evolution of both $I_1$ and $I_2$.  (In the terminology of
Geroch \cite{Geroch82}, without restrictions, we can {\it build} any
initial data set from any other initial data set.)\ \ It is worth
noting that when we restrict ourselves to spacetimes satisfying the
dominant-energy condition, if the initial data is vacuum in the sense
that the associated energy and mass-current densities (which can be
calculated using the initial-value constraint equations of general
relativity), then a possible evolution of $I$ is everywhere vacuum and
therefore is unique (up to diffeomorphisms).  More generally, however,
there will be many possible evolutions of a given initial data set.

From a possible evolution of $I$, we can calculate the least upper
bound to the lengths of causal curves having past endpoint $p$.  By
looking at this least upper bound over the set of all possible
evolutions, we can find which possible evolution gives our observer
the longest possible lifetime.  If $p$ is not within the black hole
for {\it every} possible evolution of $I$, then there is no finite
upper bound on the lifetime of our observer as there will be a
possible evolution for which there is a causal curve with past
endpoint $p$ and future endpoint on ${\cal J}^+$ (with such a curve
having infinite length).  In other words, in this case we can choose
our possible evolution so that the observer in fact was never within a
black hole.  However, if $p$ is in the subset ${\cal B}$ of $\Sigma$
consisting of all the points of $\Sigma$ that lie within the
black-hole region of each possible evolution, then an observer that
begins a journey at $p \in {\cal B}$ is {\it guaranteed} to lie within
the black-hole region independent of which possible evolution of $I$
we choose.  Note that ${\cal B}$ is a closed subset of $\Sigma$ (being
the intersection of a collection of closed sets), and that when there
is only a single possible evolution (e.g., a vacuum spacetime), then
${\cal B} = B \cap
\Sigma$.  Of course, ${\cal B}$ may be empty.  With this definition, a
question relevant to our original problem is the following.

{\it Question 1.} Given an initial data set $I$, what is the least
upper bound on the lengths of causal curves in $J^+({\cal B})$ over
the set of all possible evolutions of $I$?

Physically, this upper bound corresponds to the greatest lifetime an
observer, beginning his journey at some point in ${\cal B}$, can hope
to achieve by our (allowed) manipulations of the matter in the
spacetime.  Unfortunately, this question appears to be as difficult to
answer as proving conjectures~1 and~2.  Even the task of computing
${\cal B}$ from $I$ appears difficult as finding all possible
evolutions and their associated black-hole regions is nontrivial.  Are
there any sufficient conditions for knowing whether a given point $p
\in \Sigma$ is in ${\cal B}$?  It is well known that future trapped
surfaces must lie within the black-hole region of an asymptotically
predictable spacetime satisfying the null-convergence condition.
Further, the total trapped region ${\cal T}$ associated with an
initial data surface (or initial data set) shares this property
\cite{Wald84}.  Therefore, ${\cal T} \subset {\cal B}$.
In other words, if an observer lies in the total trapped region, it
is assured that there is no possible evolution that will allow him to
escape to infinity.  So, restricting ourselves to observers that begin
their journey from some point in ${\cal T}$, we are led to the
following question.

{\it Question 2.} Given an initial data set $I$, what is the least
upper bound on the lengths of causal curves in $J^+({\cal T})$ over
the set of all possible evolutions of $I$?

This question also appears quite difficult to answer.  However,
restricting ourselves to spherically symmetric initial data sets and
spherically symmetric possible evolutions (spherically symmetric
spacetimes), a partial answer to this last question is given by the
following theorem.

{\it Theorem 5.}  Fix a spherically symmetric initial data set $I$
with three-manifold $\Sigma \approx {\Bbb R}^3$.  Denoting the total
trapped region in $\Sigma$ by ${\cal T}$, the lengths of all causal
curves in $J^+({\cal T})$ are bounded above by $20~\max_{\cal T}(m)$
for any possible evolution satisfying the dominant-energy and
non-negative-pressures conditions \cite{ec}.

{\it Proof.}  This theorem follows from lemma~1 and the fact that $r
\le \max_{\cal T}(2m)$ on $J^+({\cal T})$, which we now establish.
As in the proof of lemma~7, we begin by noting that the boundary of
$J^+({\cal T})$ acts much like a Cauchy surface for $J^+({\cal T})$.
On $(\partial J^+({\cal T})) \setminus {\cal T}$, denote a null
generator of this null boundary by $\nu$, affinely parameterized by
$\lambda$.  Then, by the null-convergence condition (which follows
from both the dominant-energy and non-negative-pressures conditions),
we have $d^2r/d\lambda^2 \le 0$.  Therefore, as $dr/d\lambda = 0$ on
the apparent horizon ${\cal A} = \partial {\cal T}$, $dr/d\lambda \le
0$ all along $\nu$ to the future of ${\cal A}$, and so $r \le r({\cal
A})$ thereon as well.  Further, by lemma~C2, $\max_{\cal T}(r) \le
\max_{\cal T}(2m)$, where we have used the fact that $r({\cal A}) =
2m({\cal A})$.  Repeating the argument of lemma~3 in
Ref.~\cite{Burnett93} [and noting that $\nabla^a r$ cannot be
past-directed timelike on $(\partial J^+({\cal T})) \setminus {\cal
T}$], we have $r \le \max_{\cal T}(2m)$ on $J^+({\cal T})$.~$\Box$

Therefore, we have a bound on the lifetime of observers beginning
their journey in the total trapped region of a Cauchy surface in terms
of the initial data on that region.  In particular, this bound is
insensitive to the final asymptotic size of the black hole (unlike the
bounds given by theorems~1 and~2).

So, can we increase the lifetime of an observer in a black hole by
manipulating the matter?  In general, yes, since not all possible
evolutions allow for equally long causal curves.  Can we increase it
to be on the order of $M_{\text{ADM}}$?  While this remains
unanswered, for certain observers in the spherically symmetric case,
theorems~4 and~5 show that this is not always possible.  Are there any
initial data sets where adding matter cannot lengthen the lifetime of
an observer in the black hole at all?  Yes.  We state, without proof,
the following theorem giving a class of initial data (a vacuum
Schwarzschild black hole of mass $M$ surrounded by matter) showing
that the lifetime of observers in the Schwarzschild region cannot be
increased beyond the vacuum maximum $\pi M$.

{\it Theorem 6.}  Fix a spherically symmetric initial data set $I$
such that: (1) The data on its total trapped region ${\cal T}$ is
isometric to the data induced on a Cauchy surface in maximally
extended Schwarzschild spacetime of positive mass $M$ with the region
corresponding to ${\cal T}$ lying in region II of Fig.~\ref{schwarz1}
(i.e., $\nabla^a r$ is future-directed timelike everywhere inside of
${\cal T}$); (2) Outside of ${\cal T}$, $\nabla^a r$ is outgoing
spacelike.  Restricting ourselves to spacetimes satisfying the
dominant-energy and non-negative-pressures conditions \cite{ec}, the
lifetime of an observer to the future of some point $p \in {\cal T}$
under a spherically symmetric possible evolution of $I$ is no greater
than the maximum lifetime under a spherically symmetric possible
evolution with $J^+({\cal T})$ being vacuum, which is no greater than
$\pi M$.  More weakly, the least upper bound to the lengths of causal
curves in $J^+({\cal T})$ over the set of spherically symmetric
possible evolutions is $\pi M$.

\section{discussion} \label{sec:discussion}
As the proofs of conjectures~1 and~2 in the spherically symmetric case
involve a detailed use of the fields $r$ and $m$, concepts for which
we do not have an adequate generalization (at least none that are
obviously useful for our purposes), it would seem that their proof in
the general case will be a difficult task.  However, we do have a few
hints as to how one might proceed.

One possible step towards a proof of conjecture~1 is provided by
generalizing the lemma bounding the size of the spheres of symmetry
$r$ in the black hole region of a spherically symmetric spacetime
(lemma~6) along the following lines.  Consider a ``cut'' $K$ of past
null infinity ${\cal J}^-$ and consider the null boundary $\partial
J^+(K)$.  The null generators of this boundary are past complete and
have positive expansion thereon (to the past).  Therefore, the area of
a cross-section of $\partial J^+(K)$ will be decreasing to the future
and, therefore, the areas of cross-sections of $\partial J^+(K)$ in
the black-hole region will be bounded by the area associated with a
cross-section of the event horizon, which is bounded by the future
asymptotic area of the black hole's event horizon.  (This is simply a
mimicry of the proof used in lemma~6.)\ \ Even with such a result,
this is still far from showing that there are no long causal curves in
the portion of the black hole region lying to the future of ${\cal
J}^-$.

Perhaps a good place to start in the proof of conjecture~2 would be to
attempt a proof of a generalized version of theorem~5, i.e., drop the
spherically symmetry requirement and replace $\max_{\cal T}(m)$ by
some more general quantity $M_{\cal T}$ constructed on ${\cal T}$.
Although such a theorem would not place an upper bound on the lifetimes
of all observers in a black hole region, it would bound those lying to
the future of a total trapped region (which is a subset of the total
black hole region).  Further, such a theorem would be a very nice
strengthening of Penrose's 1965 theorem that shows that $J^+({\cal
T})$ cannot be future null complete \cite{Penrose65}.

We end with a few remarks and questions for consideration.  (1) Do
conjectures~1 and~2 in fact hold with $k=2\pi$ (the smallest possible
value for which they can hold as shown by our analysis of the Kerr and
dust-ball spacetimes)?  (2) Can weaker versions of conjectures~1 and~2
be proven?  For instance, a theorem that merely asserts that the
lifetimes of observers in the black-hole region will be bounded, or
merely that the lifetimes must be finite (no global upper bound),
would be of interest.  (3) If conjecture~1 or~2 is false, does a
counterexample to either conjecture exist in the literature, or are
the spacetimes providing counterexamples peculiar in some way so that
they have been missed?  (4) Can the energy conditions in theorems~1
and~2 be weakened?  In particular, it would be nice if the
non-negative-pressures condition could be replaced by a weaker
condition stating that the pressure is not ``too negative'' compared
to the energy density.  (5) As has been mentioned, Theorems~1 and~4
are slightly awkward in that they require the existence of a
geodesically complete spherically symmetric Cauchy surface so that $m$
will be non-negative everywhere as needed by lemma~1.  Can the
requirement that $m$ be non-negative in lemma~1 be dropped, thereby
allowing for improved versions of theorems~1 and~4?

\acknowledgements
I would like to thank Robert Geroch for initially directing my
attention to the possibility of proving a theorem such as
theorems~1~and~2 and for the discussions with both him and Robert Wald
regarding this work.

\appendix
\section{Bounding the area of a spherically symmetric black hole}
\label{sec:bound_area}
{\it Theorem A1.}  For a globally hyperbolic asymptotically flat
spherically symmetric spacetime satisfying the dominant-energy
condition \cite{ec} and with ${\cal J^+}$ future complete, the area
$A$ of the event horizon is always bounded above by
\begin{equation}
A \le 16\pi M_{\text{ADM}}^2.
\end{equation}
Equivalently, on the event horizon: $M_{\text{irr}} \le
M_{\text{ADM}}$; and $r \le 2 M_{\text{ADM}}$.

{\it Proof.}  The basic idea behind the proof is to construct a
``retarded-time function'' $u$ such that $u \rightarrow \infty$ as we
approach the event horizon of the black hole provided that ${\cal
J}^+$ is complete and then show that this is not this case if $r$ is
anywhere larger than $2 M_{\text{ADM}}$ on the event horizon.
Essentially, the proof uses the outgoing Eddington-Finkelstein
coordinates ($r$ and $u$, together with two angular coordinates
\cite{MTW73}).  However, we choose not to actually ``work in these
coordinates'' as it is easier to just treat $r$ and $u$ as fields and
work with their properties.

We begin by constructing a ``retarded-time function'' $u$ on
$J^-({\cal J}^+)$ as follows.  Through each point $p$ in the spacetime
with $r(p)>0$, there exists two spherically symmetric null surfaces
containing $p$.  In the region outside of the black hole, i.e.,
$J^-({\cal J}^+)$, we shall ``label'' the outgoing surfaces (by which
is meant those whose tangent vectors can be taken to be outgoing and
future-directed) by defining the scalar field $u$ to be constant on
these null surfaces and requiring that $\nabla^a u \nabla_a r
\rightarrow -1$ as we approach ${\cal J^+}$.  This choice uniquely
defines $u$ everywhere on $J^-({\cal J^+})$ (up to the addition of an
overall constant to $u$) and has the advantage that the completeness
of ${\cal J^+}$ is then equivalent to the property of $u$ assuming
all values on $J^-({\cal J}^+)$.

Set $k^a = - \nabla^a u$.  Then $k^a$ is radial, outgoing,
future-directed, null, tangent to surfaces of constant $u$, and
geodetic (as is any null vector field arising from a gradient).
Further, on $J^-({\cal J^+})$ (where $r>0$), define $l^a$ to be the
radial outgoing past-directed null vector field such that $k^a l_a =
+2$.  Then, setting $r' = k^a \nabla_a r$, we have
\begin{mathletters}
\begin{eqnarray}
l^a \nabla_a u & = & -2, \label{lu} \\
l^a \nabla_a r & = & {1 \over r'} (1-2m/r). \label{lr}
\end{eqnarray}
\end{mathletters}
Equation~(\ref{lu}) follows from the definition of $k^a$ and our
normalization requirement for $l^a$.  Equation~(\ref{lr}) follows from
the fact that $h^{ab} = k^{(a}l^{b)}$ so that $(1-2m/r) = \nabla^a r
\nabla_a r = (k^a \nabla_a r)(l^a \nabla_a r)$.  [Note that $r'(p)$ is
positive for all $p \in J^-({\cal J^+})$, as otherwise the sphere of
symmetry containing $p$ would be an outer trapped surface.  Such
surfaces cannot occur in this region provided the null-convergence
condition holds, as it does in our case.  Indeed, by the
null-convergence condition, $r'' \le 0$, and since $r'$ is unity at
${\cal J^+}$, we have $r' \ge 1$ everywhere on $J^-({\cal J^+})$.]

\begin{figure}
\input{psfig}
\centerline{\psfig{figure=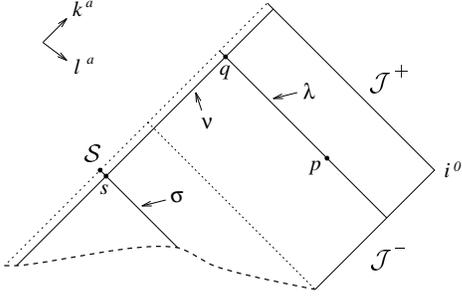}}
\medskip
\caption{The construction performed on $J^-({\cal J^+})$ used to show
that $r> r({\cal S})$ on $\Lambda = \langle\langle{\cal
J}\rangle\rangle \cap (J^-({\cal S}))^c$.  (In the case depicted here,
$\Lambda = \langle\langle{\cal J}\rangle\rangle$.)\ \ For $p \in
\Lambda$, we have $r(p) > r(q) > r(s) > r({\cal S}) - \epsilon$.
Therefore, as $r(p) > r({\cal S}) - \epsilon$ for all $\epsilon > 0$,
we have $r(p) > r({\cal S})$.  (Equality cannot be attained as
$\nabla^a r$ is outgoing spacelike on the open set $\Lambda$ .)\ \ The
dotted line denotes where the event horizon would be if it were ``all
there''.  In that case, the construction can be simplified by taking
$\nu$ to lie on the horizon, thereby avoiding the construction of
$\sigma$ and the need to take a limit as $\epsilon
\rightarrow 0$.}
\label{theorema1}
\end{figure}

Now suppose that for some sphere of symmetry ${\cal S}$ on the event
horizon $r({\cal S})> 2 M_{\text{ADM}}$.  Setting $\Lambda =
\langle\langle{\cal J}\rangle\rangle \cap (J^-({\cal S}))^c$, then $r
> r({\cal S})$ everywhere on $\Lambda$.  (See Fig.~\ref{theorema1}.)\
\ To see this, fix any point $p \in \Lambda$ and let $\lambda$ denote
the spherically symmetric ingoing null surface containing $p$.
Similarly, let $\sigma$ be the spherically symmetric ingoing null
surface containing ${\cal S}$.  Since $r$ and $u$ are continuous, for
any number $\epsilon > 0$ we can find a constant $u_0$ such that $r(s)
> r({\cal S}) -
\epsilon$ for all $s \in \sigma$ with $u(s) > u_0$.  Let $\nu$ be the
maximal integral curve of $k^a$ containing $s$.  Then, setting $q =
\nu \cap \lambda$, it follows from the fact that $\nabla^a r$ is
outgoing or past-directed (yet not parallel to $k^a$) on $J^-({\cal
J}^+)$ that $r(q) > r(s)$.  Further, choosing $s$ so that $u(s) = u(q)
> u(p)$ (i.e., so that $q$ lies to the future of $p$) we have $r(p) >
r(q)$.  Putting this all together, we have, $r(p) > r(q) > r(s) >
r({\cal S}) - \epsilon$ for all $\epsilon > 0$ and so $r > r({\cal
S})$ on $\Lambda$.  (That $r \ne r({\cal S})$ on $\Lambda$ follows
from the fact that $\nabla^a r$ is outgoing spacelike on the open set
$\Lambda$.)

Further, on $\Lambda$, we have
\begin{equation} \label{lowerbound_lr}
l^a \nabla_a r \ge 1-2 M_{\text{ADM}}/r.
\end{equation}
To see this, consider how $l^a \nabla_a r$ changes along an integral
curve of $k^a$ parameterized by $r$,
\begin{mathletters}
\begin{eqnarray}
{d \over dr} (l^a \nabla_a r)
    & = & {1 \over r'} k^b \nabla_b (l^a \nabla_a r) \\
    & = & {1 \over r'} k^b l^a \nabla_b \nabla_a r \\
    & \le & {1 \over r'} {2m \over r^2}. \label{last}
\end{eqnarray}
\end{mathletters}
The first equality follows from the definition of $r'$, the second
from the fact that $k^b \nabla_b l^a = 0$, and the third follows from
Eq.~(\ref{DDr}) and the dominant-energy condition.  Therefore,
dividing Eq.~(\ref{last}) by Eq.~(\ref{lr}) and noting that $l^a
\nabla_a r$ is positive on $\Lambda$ (as it is a subset of
$\langle\langle{\cal J}\rangle\rangle$), we have
\begin{equation} \label{bounding_lr}
(l^a \nabla_a r)^{-1} {d \over dr} (l^a \nabla_a r)
\le  {2m \over r^2} (1-2m/r)^{-1}.
\end{equation}
Using the facts that $r > 2 M_{\text{ADM}} \ge 2m$ on $\Lambda$, we
find
\begin{equation}
(l^a \nabla_a r)^{-1} {d \over dr} (l^a \nabla_a r)
\le {2 M_{\text{ADM}} \over r^2} (1-2 M_{\text{ADM}}/r)^{-1}.
\end{equation}
Integrating this and using the fact that $l^a \nabla_a r \rightarrow
1$ as $r \rightarrow \infty$ along the integral curve of $k^a$ [by
Eq.~(\ref{lr}) and the fact that $m \le M_{\text{ADM}}$ on
$\langle\langle{\cal J}\rangle\rangle$], gives
Eq.~(\ref{lowerbound_lr}).

We can now establish an upper bound on $u$ as follows.  Let $\lambda$
be a maximal integral curve of $l^a$ in $\Lambda$.  Parameterizing
this curve by $u$, we have
\begin{equation} \label{dr*du}
- {dr \over du}
     =  {(l^a \nabla_a r) \over (-l^a \nabla_a u)}
    \ge {1 \over 2}(1-2 M_{\text{ADM}}/r).
\end{equation}
Setting $r^* = r + 2 M_{\text{ADM}} \ln(r - 2 M_{\text{ADM}})$ and
integrating Eq.~(\ref{dr*du}) along the curve $\lambda$ from a point
$p$ to a point $q$ lying to the future of $p$, we have
\begin{equation} \label{bound_u}
u(q) \le u(p) + 2(r^*(p) - r^*(q)).
\end{equation}
Since $r > r({\cal S})$ on $\Lambda$, $r^*$ is bounded from below by
$r^*({\cal S})$ on $\Lambda$, which is finite as $r({\cal S}) > 2
M_{\text{ADM}}$.  Therefore, Eq.~(\ref{bound_u}) gives an upper bound
on allowed values of $u$ contradicting the completeness of ${\cal
J}^+$.  So, on the event horizon, $r$ can be no greater than $2
M_{\text{ADM}}$.~$\Box$

{\it Theorem A2.}  Fix a globally hyperbolic asymptotically flat
spherically symmetric spacetime satisfying the dominant-energy
condition \cite{ec} with ${\cal J^+}$ future complete and the event
horizon complete in the sense that a future incomplete timelike or
null geodesic in $\langle\langle{\cal J}\rangle\rangle$ can be
assigned an endpoint on the event horizon (in other words, the event
horizon is ``all there'') and the null geodesic generators of the
event horizon are future complete.  Then, $r/2$ and $m$ have the same
limit to the future on the event horizon.  (In other words, the
irreducible mass $M_{\text{irr}}$ and mass $m$ have the same future
asymptotic limits to the future on the event horizon.)

{\it Proof.}  Denote by $\lambda$ and $k^a$ the parameter and
associated tangent vector, respectively, in an affine parameterization
of a null geodesic generator $\nu$ or the event horizon.  Further, let
$l^a$ denote the radial null vector parallel transported along $\nu$
with $k^a l_a = +2$ thereon.  Denote the future asymptotic value of
$m$ on the event horizon by $M_f$.  If $r \le 2 M_f$ on all of $\nu$,
then since $2m \le r$ on the portion of $\nu$ lying to the future of
${\cal J^-}$, $r$ must have $2 M_f$ as a future limit.  So, suppose
that at some point on $\nu$ that $r > 2 M_f$.  Since $r$ is
non-decreasing to the future on $\nu$, this inequality will continue
to hold.  Then, repeating the argument that led to
Eq.~(\ref{bounding_lr}), this equation again holds with our $l^a$ on
the portion of $\nu$ lying to the future of ${\cal J^-}$.  As $m \le
M_f$ on this portion, we have
\begin{equation}
(l^a \nabla_a r)^{-1} {d \over dr} (l^a \nabla_a r) \le
{2 M_f \over r^2} (1-2 M_f/r)^{-1}.
\end{equation}
Integrating this we have
\begin{equation}
l^a \nabla_a r \le C (1 - 2 M_f/r),
\end{equation}
for some positive constant $C$.
Noting that $h^{ab} = k^{(a}l^{b)}$ on
$\nu$, so that $(k^a \nabla_a r)(l^a \nabla_a r) = 1-2m/r$, we have
\begin{equation}
{dr \over d\lambda} = k^a \nabla_a r = {1 \over (l^a \nabla_a r)}
(1-2m/r) \ge {1 \over C}.
\end{equation}
Integrating this, we see that $r \rightarrow \infty$ as $\lambda
\rightarrow \infty$.  However, this contradicts the fact that $r$ is
bounded by $2 M_{\text{ADM}}$ on the event horizon by theorem~A1.
Therefore, the event horizon cannot be complete without $r$ and $2m$
having the same future limits.~$\Box$

\section{The construction of dust-ball spacetimes}
\label{sec:constructing_dustballs}

In some detail, the construction of a dust-ball spacetime proceeds as
follows.

\begin{figure}
\input{psfig}
\centerline{\psfig{figure=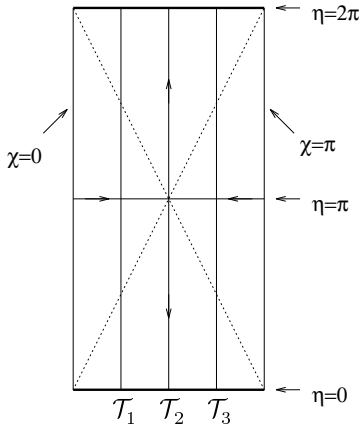}}
\medskip
\caption{A spacetime diagram representing a $k_{\text{RW}}=+1$
Robertson-Walker spacetime with dust as a source.  In this diagram,
null rays perpendicular to the spheres of symmetry have slopes of $\pm
1$.  Surfaces of constant $\eta$ are surfaces of homogeneity with
$\eta=\pi$ being the maximal hypersurface.  The two diagonal timelike
surfaces ($\eta = 2\chi$ and $\eta = 2\pi - 2\chi$) are where $r=2m$
(and $r \neq 0$).  These surfaces divide the spacetime into four
regions each on which $\nabla^a r$ points into a different quadrant.
The timelike surfaces ${\cal T}_1$, ${\cal T}_2$, and ${\cal T}_3$ are
surfaces of constant $\chi$ and are generated by the (geodetic) flow
of the dust (i.e., the integral curves of the flow are tangent to
these surfaces).  On each surface, $\chi = \chi_0$ for some constant
$\chi_0$.  For ${\cal T}_1$, $0 < \chi_0 < \pi/2$; for ${\cal T}_2$,
$\chi_0 = \pi/2$; for ${\cal T}_3$, $\pi/2 < \chi_0 < \pi$.}
\label{rwdust}
\end{figure}

The Robertson-Walker spacetimes  are spherically symmetric and can be
coordinated (everywhere excepting where $r=0$) so that the metric is
given by
\begin{equation} \label{Robertson-Walker_metric}
g_{ab} = a^2(\eta) \biglb( -(d\eta)_a(d\eta)_b + (d\chi)_a(d\chi)_b +
f_k^2(\chi)  \Omega_{ab} \bigrb),
\end{equation}
where $\Omega_{ab}$ is a unit-metric on the two-sphere,
\begin{equation}
f_{k_{\text{RW}}}(x) = \left\{
\begin{array}{ll}
\sin(x)  & \text{for } k_{\text{RW}} = +1, \\
x        & \text{for } k_{\text{RW}} = 0,  \\
\sinh(x) & \text{for } k_{\text{RW}} = -1,
\end{array}
\right.
\end{equation}
and $k_{\text{RW}}$ tells us which family we are considering.  (The
sign of $k_{\text{RW}}$ gives the sign of the intrinsic Ricci scalar
curvature of the surfaces of homogeneity.)  Surfaces of constant
$\eta$ are the surfaces of homogeneity and the integral curves of the
fluid flow lie in surfaces of constant $\chi$ and are perpendicular to
the spheres of symmetry.  Note that the allowed range for $\chi$ in
expressing the metric in this form is $\chi > 0$ in the
$k_{\text{RW}}=-1$ and $k_{\text{RW}}=0$ cases while $0 < \chi < \pi$
in the $k_{\text{RW}}=+1$ case.

From Eq.~(\ref{Robertson-Walker_metric}), the size $r$ of a sphere
of symmetry is given by
\begin{equation} \label{r_Robertson-Walker}
r(\eta,\chi) = a(\eta)f_{k_{\text{RW}}}(\chi).
\end{equation}
With dust as a source, $a(\eta)$ is given by
\begin{equation}
a(\eta) = C f_{k_{\text{RW}}}^2(\eta/2)
\end{equation}
for some constant $C$.  Using Eq.~(\ref{r_Robertson-Walker}) and
Eq.~(\ref{defm}), the mass $m$ associated with each sphere of symmetry
is calculated to be
\begin{equation} \label{m_Robertson-Walker}
m(\eta,\chi) = {C \over 2} f_{k_{\text{RW}}}^3(\chi).
\end{equation}
Notice that $m$ is constant along the integral curves of the fluid
flow, as it must be as dust has zero pressure (see Eq.~(\ref{D2m})).

To construct a dust-ball spacetime, we fix a dust-filled
Robertson-Walker spacetime as described above (i.e., fix some
$k_{\text{RW}}$ and $C$) and choose a number $\chi_0$ (within the
allowed range).  We then take the region in the Robertson-Walker
spacetime with $\chi \le \chi_0$ and attach its boundary ${\cal T}$
(being the timelike three-surface where $\chi=\chi_0$) to an
appropriate similar three-surface ${\cal T}'$ of an appropriate
extended Schwarzschild spacetime.  In order that there not be any
``surface layers'' where the spacetimes are glued together, it is
necessary that $m$ be a continuous function on the newly created
spacetime.  Therefore, the mass of the appropriate Schwarzschild
spacetime is simply $m$ evaluated on ${\cal T}$.  As the ADM mass
$M_{\text{ADM}}$ of the spacetime so constructed is the mass of the
Schwarzschild portion, we therefore have
\begin{equation} \label{dust-ball-adm-mass}
M_{\text{ADM}} = {C \over 2} f_{k_{\text{RW}}}^3(\chi_0).
\end{equation}
We shall omit the remainder of the details of this matching except to
note that, in the $k_{\text{RW}}=+1$ case, when: $\chi_0 \le \pi/2$,
${\cal T}'$ lies in regions I, II, and III and thereby ``covers up''
all of region IV in Fig.~\ref{schwarz1}; otherwise ${\cal T}'$ lies in
regions II, III, and IV and thereby cannot be reached from the
asymptotic region (region I) without entering the black hole.

In Figs.~\ref{db1} and \ref{db2} we have sketched two such dust-ball
spacetimes.  In Fig.~\ref{db1}, $\chi_0 < \pi/3$, while in
Fig.~\ref{db2}, $\chi_0 > 2\pi/3$.  In each, we have divided the
spacetime into a number of regions according to the following scheme.
On I, $\text{I}'$, $\text{I}''$, and $\text{I}'''$, $\nabla^a r$ is
outgoing and: I lies to the future of ${\cal J^-}$ and the past of
${\cal J^+}$ (i.e, I is the asymptotic region); $\text{I}'$ is in the
future of ${\cal J^-}$ but not in the past of ${\cal J^+}$;
$\text{I}''$ lies past of ${\cal J^+}$ but not in the future of ${\cal
I^-}$; $\text{I}'''$ lies neither in the future of ${\cal J^-}$ nor in
the past of ${\cal J^+}$.  On II and $\text{II}'$, $\nabla^a r$ is
future-directed and II lies to the future of ${\cal J^-}$ while
$\text{II}'$ does not.  On III and $\text{III}'$, $\nabla^a r$ is
past-directed and III lies to the past of ${\cal J^+}$ while
$\text{III}'$ does not. On IV, $\nabla^a r$ is ingoing.  For the
dust-ball spacetimes, regions I, II, and III always exist while:
$\text{I}'$ and $\text{I}''$ exist only for $\chi_0 < 2\pi/3$;
$\text{I}'''$ exists only for $\chi_0 > \pi/3$; $\text{II}'$,
$\text{III}'$, and IV exist only for $\chi_0 > \pi/2$.

Note that the outer surface of the dust ball crosses the event horizon
of the black hole where the surface $\chi=\chi_0$ intersects the
surface $\eta = 2\pi - 2\chi$, i.e., at $\eta = \eta_0 = 2\pi - 2
\chi_0$.  (See Figs.~\ref{rwdust}, \ref{db1}, and \ref{db2}.)\ \
Therefore, on the event horizon in the dust-filled portion of the
spacetime we have $\eta = \eta_0 + (\chi-\chi_0) = (2\pi - 3\chi_0) +
\chi$.  In particular, in the $k_{\text{RW}}=+1$ case, we see that the
black hole is eternal, in the sense that every Cauchy surface will
intersect the black-hole region, if $\chi_0 \ge 2\pi/3$.

\section{Two lemmas bounding $\lowercase{r}$} \label{sec:bounding_r}

{\it Lemma C1.} Fix globally hyperbolic spherically symmetric spacetime
satisfying the non-negative-pressures condition \cite{ec}.  Fix any
Cauchy surface surface $\Sigma$ therein and let $P$ denote the subset
of $\Sigma$ on which $\nabla^a r$ is past-directed timelike,
past-directed null, or zero.  Then, for a point $p \in D^+(\Sigma)$ at
which $\nabla^a r$ is past-directed timelike, past-directed null, or
zero, we have
\begin{equation}
r(p) \le 2m(p) \le \sup_P(2m).
\end{equation}

{\it Proof.}  The first inequality is immediate from the definition of
$m$ while the proof of the second is obtained by repeating the
argument given in the latter portion of the proof of lemma~3 in
Ref.~\cite{Burnett93} and noting that the theorem's requirement that
the dominant-energy condition hold can be dropped.  (That is, as the
null-convergence condition follows from the non-negative-pressures
condition and it can be shown that the existence of a timelike vector
$t^a$ for which the last term in Eq.~(2.16) in Ref.~\cite{Burnett93}
is non-negative follows from the null-convergence condition, the
requirement that the dominant-energy condition hold was
superfluous.)~$\Box$

{\it Lemma C2.} Fix a globally hyperbolic spherically symmetric
spacetime and a spherically symmetric Cauchy surface $\Sigma$ therein.
For a spherically symmetric compact subset $C$ of $\Sigma$, define
$\partial^+C$ to be those components of $\partial C$ for which
$\nabla^a r$ is spacelike and $v^a \nabla_a r > 0$ for vectors $v^a$
on $\partial C$ (tangent to $\Sigma$) and pointing out of $C$.  Then,
we have
\begin{equation} \label{bound_r_on_C}
\max_C(r) \le \max \biglb( \max_{\partial^+ C}(r), \max_C(2m) \bigrb).
\end{equation}

{\it Proof.}  Let $p$ be any point where $r$ reaches its maximum value
on $C$.  If $\nabla^a r$ is non-spacelike at $p$, then $\max_C (r) =
r(p) \le 2m(p) \le \max_C(2m)$, where the first inequality follows
from the definition of $m$ given by Eq.~(\ref{defm}).  Otherwise, if
$\nabla^a r$ is spacelike at $p$, then $p$ must be on $\partial^+ C$
for otherwise if $p$ is in the interior of $C$ or on any other part of
the boundary, there exist vectors $v^a$ (either in $C$ or pointing
into $C$) such that $v^a \nabla_a r > 0$ thereby violating the
maximality of $r$ at $p$ on $C$.  Therefore, in this case $\max_C(r) =
r(p) =
\max_{\partial^+C}(r)$.  Combining these two cases,
Eq.~(\ref{bound_r_on_C}) follows.~$\Box$


\begin{references}
\bibitem{af_defs}
A null infinity associated with an asymptotically flat spacetime is
denoted here by ${\cal J}$, which is the disjoint union of a past part
${\cal J}^-$ (past null-infinity) and a future part ${\cal J}^+$
(future null-infinity).  Spatial infinity is denoted by $i^0$.  (See
Ref.~\cite{Wald84} for a discussion of these notions.)
\bibitem{ec}
A spacetime with Ricci tensor $R_{ab}$ and Einstein tensor $G_{ab}$ is
said to satisfy: (1) the null-convergence condition (also known as the
null energy condition) if $R_{ab}k^a k^b \ge 0$ (equivalently
$G_{ab}k^ak^b \ge 0$) for all null $k^a$; (2) the timelike-convergence
condition (also known as the strong energy condition) if $R_{ab}t^at^b
\ge 0$ for all timelike $t^a$; (3) the non-negative-energy condition
(also known as the weak energy condition) if $G_{ab}t^at^b \ge 0$ for
all timelike $t^a$; (4) the dominant-energy condition if $G_{ab}t^a
u^b \ge 0$ for all future-directed timelike $t^a$ and $u^b$; (5) the
non-negative-pressures condition if $G_{ab}x^a x^b \ge 0$ for all
spacelike $x^a$.
\end{references}
\end{document}